\documentclass[smallextended]{svjour3}       
\smartqed  

\usepackage{graphicx}
\graphicspath{{./fig/}}
\DeclareGraphicsExtensions{.pdf,.jpeg,.png,.eps}
\usepackage{amsmath}
\usepackage{amssymb}

\usepackage{amsthm}
\usepackage{subfigure}
\usepackage{amsfonts}
\usepackage{multirow}
\usepackage{algorithm}
\usepackage{algorithmic}
\usepackage{url}
\usepackage{color}
\usepackage{epstopdf}
\newtheorem*{def1}{Definition}{\bfseries}{\itshape}
\newtheorem*{obj}{Objective}{\bfseries}{\itshape}
\newtheorem*{mod1}{Scale relationship model}{\bfseries}{\itshape}
\usepackage{natbib}

\begin{document}

\title{Multiscale Event Detection in Social Media}

\author{Xiaowen Dong          \and
        Dimitrios Mavroeidis      \and
        Francesco Calabrese    \and
        Pascal Frossard
}

\institute{X. Dong \at
              MIT Media Lab, Cambridge, MA, United States\\
              \email{xdong@mit.edu}
           \and
           D. Mavroeidis \at
              Philips Research, Eindhoven, Netherlands\\
              \email{dimitrios.mavroeidis@philips.com}
           \and
           F. Calabrese \at
              IBM Research, Dublin, Ireland\\
              \email{fcalabre@ie.ibm.com}
           \and
           P. Frossard \at
              \'{E}cole Polytechnique F\'{e}d\'{e}rale de Lausanne (EPFL), Lausanne, Switzerland\\
              \email{pascal.frossard@epfl.ch}
}

\date{Received: date / Accepted: date}

\maketitle

\begin{abstract}
Event detection has been one of the most important research topics in social media analysis. Most of the traditional approaches detect events based on fixed temporal and spatial resolutions, while in reality events of different scales usually occur simultaneously, namely, they span different intervals in time and space. In this paper, we propose a novel approach towards multiscale event detection using social media data, which takes into account different temporal and spatial scales of events in the data. Specifically, we explore the properties of the wavelet transform, which is a well-developed multiscale transform in signal processing, to enable automatic handling of the interaction between temporal and spatial scales. We then propose a novel algorithm to compute a data similarity graph at appropriate scales and detect events of different scales simultaneously by a single graph-based clustering process. Furthermore, we present spatiotemporal statistical analysis of the noisy information present in the data stream, which allows us to define a novel term-filtering procedure for the proposed event detection algorithm and helps us study its behavior using simulated noisy data. Experimental results on both synthetically generated data and real world data collected from Twitter demonstrate the meaningfulness and effectiveness of the proposed approach. Our framework further extends to numerous application domains that involve multiscale and multiresolution data analysis.

\keywords{Multiscale event detection \and Spatiotemporal analysis \and Wavelet decomposition \and Modularity-based clustering}
\end{abstract}

\section{Introduction}
\label{intro}
The last decade has seen rapid development of online social networks and social media platforms, which leads to an explosion of user-generated data posted on the Internet. The huge amount of such data enables the study of many research problems, and event detection is certainly one of the most popular and important topics in this novel research area. Social media platforms present several advantages for event detection. First, due to the real-time nature of online social services, the public awareness of real world happenings could be raised in a much quicker fashion than with the traditional media. Second, due to the large amount of users posting content online, more complete pictures of the real world events with descriptions from different angles are offered with fast and large-scale coverage. These advantages have attracted a significant amount of interest from the data mining communities in event detection problems. For instance, the MediaEval Workshop has open research task dedicated to event detection \citep{Reuter13}, and numerous event detection approaches have been proposed recently in the literature \citep{Sayyadi09,Becker09,Aggarwal12}.

\begin{figure}[t]
	\begin{center}
		\begin{tabular}{cc}
			~\includegraphics[width=0.45\textwidth]{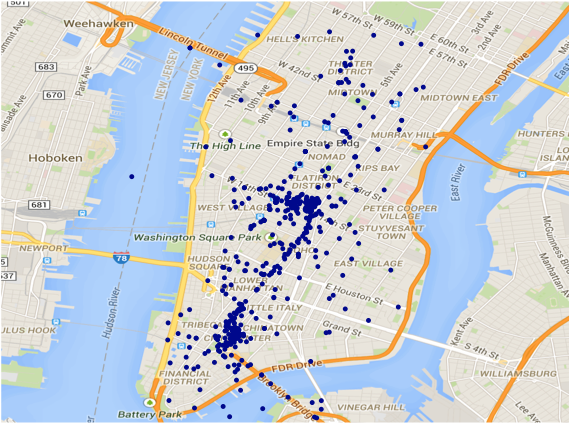}~ & ~\includegraphics[width=0.45\textwidth]{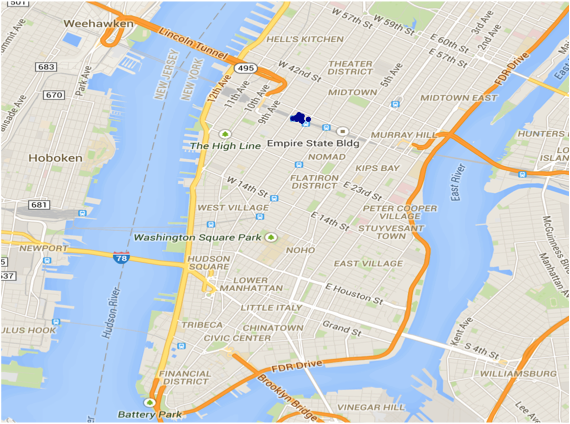}~\\
			~(a)~ & ~(b)~\\
		\end{tabular}
	\end{center}
	\vspace{-0.4cm}
	\caption{Two events in New York City that has been discussed on Twitter: (a) Occupy Wall Street protests. (b) Katy Perry's concert at the Madison Square Garden. Each dot on the map represents a tweet. While discussions about the first event span the middle and lower Manhattan area for more than three hours, the discussions about the second are concentrated near the concert venue for less than an hour.}
	\label{fig:event_examples}
\end{figure}

Events in social media platforms can be loosely defined as real world happenings that occur within similar time periods and geographical locations, and that have been mentioned by the online users in the forms of images, videos or texts. Different types of events are usually of different temporal and spatial \textit{scales} or \textit{resolutions}\footnote{Throughout the paper, we use ``scales'' and ``resolutions'' interchangeably.}, meaning that they span different \textit{intervals} in time and space. For example, discussions about the London 2012 Summer Olympic Games would span a temporal period of nearly one month and a spatial area of all over the world, while those regarding the 2012 concert of The Stone Roses in the Phoenix Park in Dublin may concentrate only on the date and at the location of the concert. Similarly, Fig.~\ref{fig:event_examples} illustrates discussions on Twitter about two events of different spatiotemporal scales in New York City. In the designs of event detection algorithms, it is thus important to take into account the different temporal and spatial scales of various kinds of events. This is challenging in the sense that: (i) Event detection approaches usually rely on classification or clustering algorithms with fixed temporal and spatial resolutions; This results in the detected events being of similar scales; (ii) It is not yet clear how multiple resolutions in time and in space interact with each other so that they can be analyzed simultaneously, even if it is relatively easier to take into account multiple resolutions in only one of these two dimensions; (iii) Data streams from social media platforms usually contain much noisy information irrelevant to the events of interest. It is thus important to understand how to attenuate the influence of the noise on detecting events of different scales. Efficient and robust multiscale event detection for solving the above challenges is exactly the objective of the present paper.

In this paper, we first introduce a baseline approach that detects events that are of similar scales and localized in both time and space, which serves as a first step towards the understanding of multiscale event detection.
We then propose a novel approach towards the detection of events that are of different scales and localized either in time or in space but not necessarily in both simultaneously. To this end, we study the relationship between scales in the two dimensions and explore the properties of the wavelet transform
to automatically and explicitly handle the interaction between different scales in time and space simultaneously. We propose an algorithm to compute a data similarity graph at appropriate scales, based on which we perform a graph-based clustering process to detect events of different spatiotemporal scales. Furthermore, we present spatiotemporal analysis of the distribution of noisy information in data streams, especially using notions from spatial statistics, which allows us to define a novel term-filtering procedure for the proposed multiscale event detection algorithm, and helps us study the behavior of the two approaches in this paper using simulated noisy data.

We compare the proposed multiscale event detection approach with the baseline approach on both synthetically generated data and real world data collected from Twitter. We show experimentally that the proposed approach can effectively detect events of different temporal and spatial scales. On the one hand, we believe that the modeling of the relationship and interaction between temporal and spatial scales and the detection of multiscale events provide new insights into the task of event detection with social media data. On the other hand, the proposed framework can be further generalized to other application domains that involve multiscale or multiresolution data analysis.


\section{Spatiotemporal detection of events}
\label{sec:formulation}
In this paper, we define an ``event'' in social media as follows.
\begin{def1}
Events in social media are real world happenings that are reflected by data that are concentrated either in both time and space, or in at least one of the two dimensions.
\end{def1}
\noindent Events defined as above are usually of different temporal and spatial scales, namely, they span different intervals in time and space. In addition, there exist data that do not contain any information about ongoing events. In the case of Twitter, such examples can be tweets that are like: ``At work'', or ``It feels great to be home...''. When non-informative tweets constitute a large part of the input data, the event-relevant tweets could however be buried in noise. It becomes very difficult in this case to identify the information of interest. In this paper, we focus on the Twitter data streams and consider the following objective.
\begin{obj}
Consider a Twitter data stream that contains temporal, spatial and text information. Our goal is to design event detection approaches that (i) are able to identify events that appear at multiple spatiotemporal scales, namely, events that affect or take place in different temporal and spatial intervals, and (ii) are robust against the ambiguous and noisy information present in the data.
\end{obj}
In this paper, we cast event detection as a graph-based clustering problem, where the vertices of the graph represent the tweets, and the edges reflect their similarities. The goal is to group similar tweets into the same cluster such that they correspond to a real world event. The clustering algorithm utilizes a similarity measure between tweets that takes into account the temporal, spatial, and textual features of a tweet. Intuitively, two tweets that are generated by users that are participating in the same event should share a number of common terms and be closely located in time and/or space. In this paper, we compare two different ways of measuring similarity between tweets, the first a baseline approach based on spatiotemporal constraints and the second a novel wavelet-based scheme. Then, in order to effectively handle the noisy information, we study the spatiotemporal distribution of the noise in the Twitter data, especially using a homogeneous Poisson process as a statistical model in our analysis. This is helpful to analyze the behavior of the baseline and the proposed event detection algorithms.

\section{Local event detection via spatiotemporal constraints}
\label{sec:led}
Events defined as in the previous section can have different localization behavior in time and space. When the events are localized in both dimensions, event detection can be effectively implemented by imposing spatiotemporal constraints on the data. In this section, we first describe a baseline approach for detecting events that are localized both in time and space, which serves as a first step towards the understanding of multiscale event detection presented later. We formulate a clustering problem, where we wish to group together the tweets that correspond to the same real world event. The similarity measure between different tweets is thus important. In our baseline event detection approach, we measure the similarity between every pair of tweets $t_i$ and $t_j$ as:
\begin{equation}
S_1(t_i,t_j)=\left\{
\begin{array}{ll}
s_\text{tf-idf}(t_i,t_j) \quad & \text{if} \quad t(t_i,t_j) \leq T_t~\text{and}~d(t_i,t_j) \leq T_d, \\
0 & \text{otherwise}.
\end{array} \right.
\label{eq:local}
\end{equation}
where $t(t_i,t_j)$ and $d(t_i,t_j)$ are the temporal difference in minutes and the spatial distance in meters, respectively, between $t_i$ and $t_j$. The thresholds $T_t$ and $T_d$ enforce the locality of the events and impose strict spatiotemporal constraints. Under such constraints, two tweets $t_i$ and $t_j$ that have a reasonably high text similarity tend to refer to the same event in real world. The function $s_\text{tf-idf}(t_i,t_j)$ represents the text similarity of $t_i$ and $t_j$ in terms of the cosine angle between the vector representations of the two tweets using the \textit{term frequency-inverse document frequency (tf-idf)} weighting scheme \citep{Manning08}. 

Given $S_1(t_i,t_j)$ as the pairwise similarity between tweets, we can create an undirected and weighted graph with adjacency matrix $W_1$:
\begin{equation}
W_1(i,j)=\left\{
\begin{array}{ll}
S_1(t_i,t_j) \quad & \text{if} \quad i \neq j, \\
0 & \text{if} \quad i=j,
\end{array} \right.
\label{eq:localgraph}
\end{equation}
where the vertices represent tweets and the edges (along with the associated weights) are defined by $S_1(t_i,t_j)$. By partitioning the vertices of the graph into disjoint clusters, each cluster is then expected to contain tweets that are likely to correspond to the same event. Furthermore, due to the constraints introduced in Eq.~(\ref{eq:local}), these events are localized in both time and space. In this paper, we perform graph-based clustering using the Louvain method \citep{Blondel08}. This is a greedy optimization method that first find small communities in a local way by maximizing the modularity function \citep{Newman06}, before repeating the same procedure by considering the communities found in the previous step as vertices in a new graph, until a maximum of modularity is attained\footnote{Since we are interested in local clusters, we apply the non-recursive version of the Louvain method which stops after the first iteration.}. The Louvain method is suitable for our purpose of event detection because of the following advantages: (i) Unlike most of the clustering methods, it does not require prior knowledge about the number of clusters; This is important because we usually do not know the number of events a priori. (ii) Unlike the popular approach based on normalized graph cut (such as spectral clustering \citep{Luxburg07}), it does not necessarily favor a balanced clustering; This enables the detection of small-scale clusters together with some relatively larger ones. (iii) It is also computationally very efficient when applied to large scale networks. Specifically, the complexity of the greedy implementation in \citet{Blondel08} is empirically observed to be close to $\mathcal{O}(n~\text{log}~n)$ where $n$ is the number of the vertices in the graph.

\begin{figure}[t]
	\begin{center}
		\begin{tabular}{cc}
			~\includegraphics[width=0.45\textwidth]{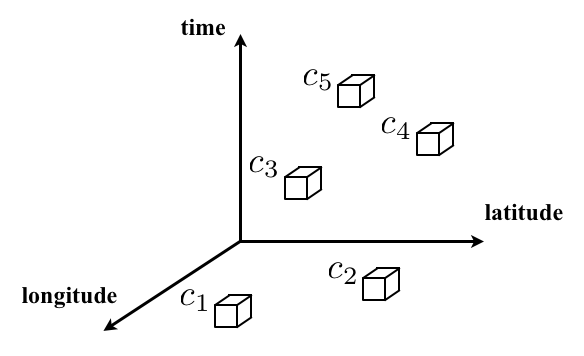}~ & ~\includegraphics[width=0.45\textwidth]{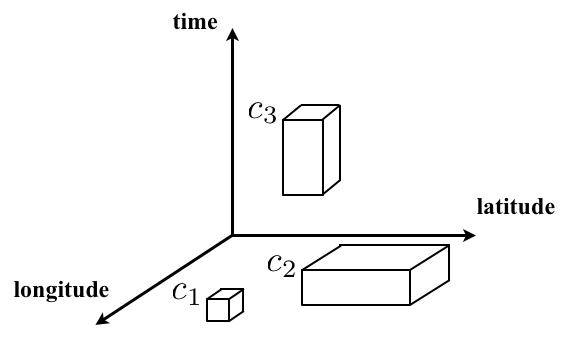}~\\
			~(a)~ & ~(b)~\\
		\end{tabular}
	\end{center}
	\vspace{-0.4cm}
	\caption{(a) Events of similar scales and are localized in both time and space. (b) Events of different scales and are not necessarily localized in both time and space.}
	\label{fig:eventscale}
\end{figure}

The graph-based clustering approach described above outputs a set of clusters that correspond to events localized in both time and space. This can be illustrated by Fig.~\ref{fig:eventscale}(a), where each cluster corresponds to a particular time-space ``cube''. After clustering, we apply simple post-processing steps to identify those clusters that are likely to correspond to meaningful events in real world. For example, we consider that a meaningful event should be observed by a sufficient number of users with sufficient information reflected on Twitter. Therefore, we consider a cluster as a local event if and only if the number of tweets and distinct Twitter users within the cluster are above certain thresholds (see Sect.~\ref{sec:realresults} for the implementation details of these post-processing steps). The algorithm for local event detection is summarized in Algorithm \ref{alg:led}.

Obviously, the choices for the values of thresholds $T_t$ and $T_d$ in Eq.~(\ref{eq:local}) are critical in \textbf{LED}. Without prior information we may choose them such that they correspond to the expected temporal and spatial spans of events to be discovered. By setting $T_t$ and $T_d$ appropriately, the algorithm would then be efficient at detecting events that are of similar scales and that are sufficiently concentrated in both time and space. For events of different scales, however, setting the thresholds too low might break down some event clusters while setting them too high would generally lead to a higher amount of noisy information in other clusters\footnote{One may think of applying \textbf{LED} with small values for $T_t$ and $T_d$ before grouping similar clusters together using a second clustering step. In fact, the second and further iterations of the Louvain method already offers such a grouping. Alternatively, a hierarchical clustering algorithm can be applied to the clusters obtained by \textbf{LED}. However, such further grouping process does not usually lead to a clear interpretation in terms of the spatiotemporal scales of the resulting event clusters, and it is often difficult to decide when to stop the recursive process and output the eventual clusters.} (as we will see in the experimental sections). In this case, one needs to implement more complex detection schemes to identify events that appear at multiple spatiotemporal scales. Hence, we introduce in the next section our novel wavelet-based method for multiscale event detection.

\begin{algorithm}[h]
\caption{Local Event Detection via Locality Constraints \textbf{(LED)}}
\begin{algorithmic}[1]

\STATE
\textbf{Input:} \\
$\mathcal{T}$: a set of tweets with temporal, spatial, and text information \\
$T_t$: temporal threshold \\
$T_d$: spatial threshold \\

\STATE
Compute the pairwise similarities $S_1(t_i,t_j)$ between tweets in $\mathcal{T}$ using Eq.~(\ref{eq:local}), and the adjacency matrix $W_1$ using Eq.~(\ref{eq:localgraph}).

\STATE
Apply the non-recursive Louvain method to $W_1$, and retain the meaningful clusters $\{c_i\}_{i=1}^m$ after post-processing steps.

%

\STATE
\textbf{Output:} \\
$\{c_i\}_{i=1}^m$: clusters that correspond to events that are localized in both time and space.\\

\end{algorithmic}
\label{alg:led}
\end{algorithm}

\section{Multiscale event detection using wavelets}
\label{sec:med}
In this section, we propose a novel algorithm for multiscale event detection. Specifically, we first introduce a new model of the relationship and interaction between the temporal and spatial scales. We then propose a wavelet-based scheme for computing the pairwise multiscale similarities between tweets.


\subsection{Relationship model between temporal and spatial scales}
The fundamental question in designing approaches towards multiscale event detection resides in properly handling events that are of different scales and do not have simultaneous temporal and spatial localization. An illustration is shown in Fig.~\ref{fig:eventscale}(b), where three events are represented by rectangular cuboids that span different time and space intervals. Two of them are only concentrated in one dimension but spread in the other one. In such cases, we need to compute a similarity score $S_2(t_i,t_j)$ between pairs of tweets $t_i$ and $t_j$ that carefully considers the temporal and spatial scales of different events. We shall relax the strict constraints in both temporal and spatial dimensions as defined in Eq.~(\ref{eq:local}), so that $S_2(t_i,t_j)$ is computed at appropriate scales that actually correspond to the span of the underlying events. To this end,  
we propose in this paper to model the relationship and interaction between the temporal and spatial scales as follows.
\begin{mod1}
When two tweets $t_i$ and $t_j$ share common terms and are close in space, we could tolerate a coarser temporal resolution in computing $S_2(t_i,t_j)$. Vice versa, when they are close in time, we could tolerate a coarser spatial resolution.
\label{mod:timespace}
\end{mod1}
\noindent Our scale relationship model essentially says that, for two tweets $t_i$ and $t_j$ to be considered similar, they should be similar at a fine resolution in at least one of the temporal or spatial dimensions, but not necessarily in both simutaneousy. It thus represents a tradeoff between time and space in the detection of events of different spatiotemporal scales. This matches the observation 
that real world events often happen within a small geographical area but could span longer time intervals (such as a protest at a certain location in a city), or they take place only within short time intervals but could spread a larger geographical area (such as a brief power outage across different areas of a city). Therefore, based on the proposed model, we can relax the strict constraints defined in Eq.~(\ref{eq:local}) in event detection.

In order to do so, however, we do not compare two tweets $t_i$ and $t_j$ with large temporal or spatial distances by simply choosing higher thresholds $T_t$ and $T_d$, since this would suffer from text ambiguity generally present in the Twitter data stream (the same word having different meanings depending on context). We do not either incorporate directly the exact temporal and spatial distances between them into the computation of the similarity metric $S_2(t_i,t_j)$, since this might lead to domination of one scale to the other. These limitations motivate us to propose a more detailed analysis model, that is, instead of considering the temporal and spatial information of each tweet as a whole, we now analyze spatiotemporal patterns of the terms (or keywords) contained in each tweet. More specifically, to compare two tweets $t_i$ and $t_j$, we propose to look at the similarity between the time series of the number of occurrences of the common terms shared by them (the occurrence is evaluated in terms of in how many tweets these terms appear). On the one hand, this enables us to study the interaction between the temporal and spatial scales when computing the similarity between keyword time series. On the other hand, this does not affect the clustering-based event detection framework, as similarities between tweets would eventually be computed based on similarities between time series of the common terms shared by them.

\begin{figure*}[t]
	\begin{center}
		\begin{tabular}{cc}
			~\includegraphics[width=0.75\textwidth]{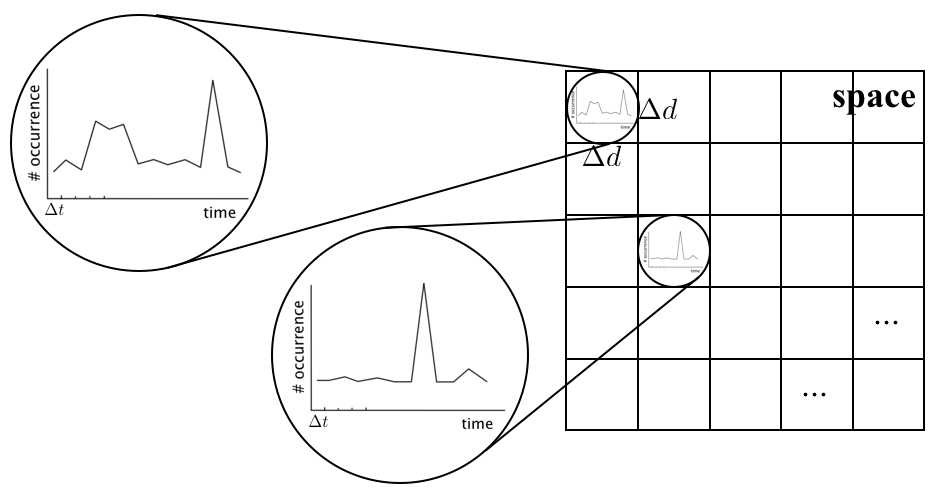}~\\
		\end{tabular}
	\end{center}
	\vspace{-0.4cm}
	\caption{Two time series of the number of occurrences of a certain term (computed using the temporal resolution $\Delta t$) within two different geographical cells. These geographical cells are defined by discretizing the geographical area using the spatial resolution $\Delta d$.}
	\label{fig:scalesinteraction}
\end{figure*}

We build the time series of keywords as follows. We start with initial temporal resolution $\Delta t$ and spatial resolution $\Delta d$. Next, for each term shared by $t_i$ and $t_j$, we compute using the temporal resolution $\Delta t$ two time series of its number of occurrences, that are based on data corresponding to the two geographical cells to which $t_i$ and $t_j$ belong. These geographical cells are defined by discretizing the geographical area using the spatial resolution $\Delta d$. The keyword time series are illustrated in Fig.~\ref{fig:scalesinteraction}.

\subsection{Wavelet-based similarity computation}
\begin{figure*}[t]
	\begin{center}
		\begin{tabular}{cc}
			~\includegraphics[width=0.75\textwidth]{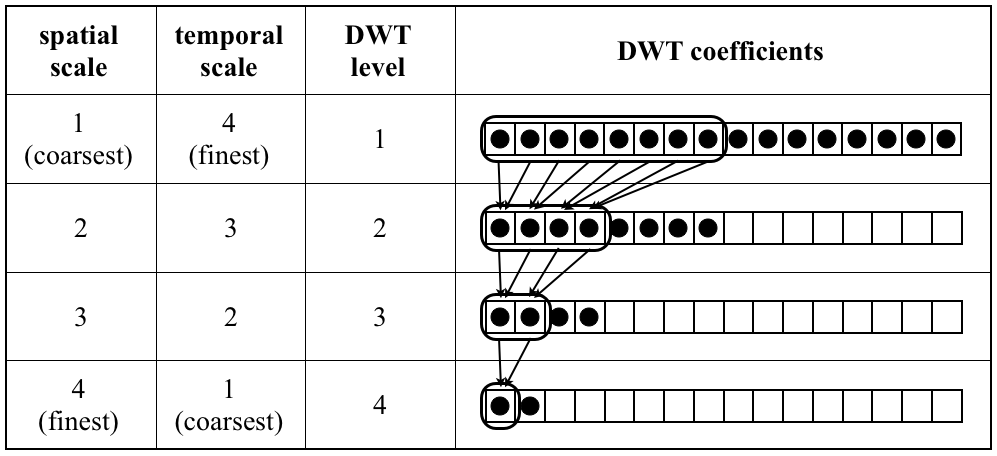}~\\
		\end{tabular}
	\end{center}
	\vspace{-0.4cm}
	\caption{An illustration of wavelet-based similarity computation for time series of length 16 (zero padding can be applied when needed if the length of time series is not a power of 2). The dots indicate the DWT coefficients at the corresponding level that are used for computing correlation, while the dots highlighted by the rounded rectangle correspond to approximation coefficients at each level. The approximation coefficients at DWT level $k$ correspond to a time series generated by aggregating every $2^k$ entries in the original time series (up to a constant).}
	\label{fig:dwt}
\end{figure*}

We now propose to use a wavelet-based method to measure similarities between time series of keywords. Similarity between time series are often measured by the correlation of their coefficients under the wavelet transform \citep{Daubechies92}, which is a well-developed tool in signal processing that leads to a multiresolution representation of the signals. In this paper, we consider the discrete wavelet transform (DWT) using the Haar wavelet, since it provides a natural way to handle different temporal scales as required in our approach. Specifically, due to the properties of the Haar wavelet, the approximation coefficients of DWT at different levels naturally correspond to aggregating the time series from fine scales (starting with the initial temporal resolution) into coarse scales, each time by a factor of two. Therefore, to evaluate the similarity of the time series at a certain temporal scale, we only need to measure the correlation between a specific set of the DWT coefficients at the corresponding level (see Fig.~\ref{fig:dwt} for an illustration).

Our key idea is then to evaluate the similarity between the two time series shown in Fig.~\ref{fig:scalesinteraction} at a properly chosen temporal scale, which is in turn determined by the spatial distance between the two geographical cells. More specifically, we introduce a number of predefined spatial scales for the spatial distance. Then, if the spatial scale is coarse, which means that $t_i$ and $t_j$ are distant, then we require the time series to be compared at a finer temporal scale (the finest temporal scale being the initial temporal resolution); Alternatively, if the spatial scale is fine, which means that $t_i$ and $t_j$ are close, then the time series could be compared at a coarser temporal scale. Given the number of spatial scales specified by the parameter $n_\text{scale}$, we define $n_\text{scale}$ distance ranges using logarithmical equispacing between the minimum and maximum distances between two distinct geographical cells (measured based on the center of the cells), which correspond to these spatial scales\footnote{When two tweets come from the same geographical cell, they would share the same time series for any common term. In this case, the correlation of DWT coefficients would always be 1 regardless of the level at which we compute the transform (or the temporal scale). This special case can be interpreted as only keeping the spatial constraint in \textbf{LED} but relaxing the temporal constraint.}. According to the scale relationship model, the temporal scale $\mathcal{S}_t$ is then selected inversely according to the spatial scale:
\begin{equation}
\mathcal{S}_t = n_\text{scale}+1 - \mathcal{S}_s.
\label{eq:computescale}
\end{equation}
For instance, if we choose to have $n_\text{scale}=4$ spatial scales $\mathcal{S}_s \in \{1,2,3,4\}$, 1 being the coarsest and 4 the finest, then we would have $\mathcal{S}_t=4,3,2,1$, respectively, that represent from the finest to the coarsest temporal scale. This in turn means that we compute the DWT at levels from 1 to 4, respectively. This procedure is illustrated in Fig.~\ref{fig:dwt}.

We can now define a new similarity metric between two tweets $t_i$ and $t_j$ as follows:
\begin{equation}
S_2(t_i,t_j) = s_\text{tf-idf}(t_i,t_j) \times s_{st}(t_i,t_j), \\
\label{eq:ms}
\end{equation}
where $s_\text{tf-idf}(t_i,t_j)$ is the text similarity of $t_i$ and $t_j$ defined as in Eq.~(\ref{eq:local}). For each term shared by $t_i$ and $t_j$, we can compute a similarity of the corresponding time series; $s_{st}(t_i,t_j)$ is then defined as the maximum such similarity among all the terms shared by $t_i$ and $t_j$. The reasons why we choose the maximum similarity are as follows. First, social media platforms that are ideal for event detection usually contain short textual data where two pieces of text, if corresponding to the same event, would share only a few but informative common terms, such as hashtags in Twitter or tags in Youtube or Flickr. Second, in Twitter specifically, although many tweets may share the same popular term, it is less often that there would be a high similarity between the two keyword time series in terms of their spatiotemporal patterns, especially at fine temporal scales, after a term-filtering procedure which we propose in the next section that removes the ``noisy'' terms that generally spread in time and space. We thus consider high similarity between time series as a strong indicator that $t_i$ and $t_j$ may be related to the same event. Taking the maximum instead of the average similarity helps us preserve such information and promote a higher recall metric (retrieval of positive links between tweets) that we favor. In Eq.~(\ref{eq:ms}), we consider the overall similarity between two tweets as a product of their text similarity ($s_\text{tf-idf}(t_i,t_j)$) and the similarity of spatiotemporal patterns of the terms shared by them ($s_{st}(t_i,t_j)$). This leads to an interesting comparison between $\textbf{LED}$ and $\textbf{MED}$: Both approaches only consider the text similarity $s_\text{tf-idf}(t_i,t_j)$ that is meaningful in event detection; However, while the former relies on fixed temporal and spatial constraints on $t_i$ and $t_j$, the latter looks at similar spatiotemporal patterns of the common terms, thus offers more flexibility for events of different scales. Finally, we can use our new similarity metric to construct an undirected and weighted graph $W_2$:
\begin{equation}
W_2(i,j)=\left\{
\begin{array}{ll}
S_2(t_i,t_j) \quad & \text{if} \quad i \neq j, \\
0 & \text{if} \quad i=j,
\end{array} \right.
\label{eq:msgraph}
\end{equation}
Based on this similarity graph, we can again apply the Louvain method to detect event clusters. The complete algorithm for the proposed multiscale event detection approach is summarized in Algorithm \ref{alg:med}.

\begin{algorithm}[h]
\caption{Multiscale Event Detection using Wavelets \textbf{(MED)}}
\begin{algorithmic}[1]

\STATE
\textbf{Input:} \\
$\mathcal{T}$: a set of tweets with temporal, spatial, and text information \\
$\Delta t$: initial temporal resolution \\
$\Delta d$: initial spatial resolution \\
$n_\text{scale}$: number of spatial scales \\

\STATE
For every pair of tweets $t_i$ and $t_j$ in $\mathcal{T}$, extract the common terms $\{w_i\}_{i=1}^k$.

\STATE
For each $w_i$, compute using $\Delta t$ the time series of its number of occurrences, that are based on data corresponding to the two geographical cells (defined using $\Delta d$) to which $t_i$ and $t_j$ belong.

\STATE
Determine using Eq.~(\ref{eq:computescale}) the temporal scale $\mathcal{S}_t$ using the spatial scale $\mathcal{S}_s$ to which the distance between the two geographical cells corresponds.

\STATE
Apply DWT to the two time series, and compute the similarity between them as the correlation between a specific set of DWT coefficients at the level corresponding to $\mathcal{S}_s$.

\STATE
Compute $s_{st}(t_i,t_j)$ as the maximum time series similarity among $\{w_i\}_{i=1}^k$. Compute $S_2(t_i,t_j)$ using Eq.~(\ref{eq:ms}), and the adjacency matrix $W_2$ using Eq.~(\ref{eq:msgraph}).

\STATE
Apply the non-recursive Louvain method to $W_2$, and retain the meaningful clusters $\{c_i\}_{i=1}^m$ after post-processing steps.

\STATE
\textbf{Output:} \\
$\{c_i\}_{i=1}^m$: clusters that correspond to events of different temporal and spatial scales.\\

\end{algorithmic}
\label{alg:med}
\end{algorithm}

There are a number of parameters in our multiscale event detection approach. First, the initial resolution parameters $\Delta t$ and $\Delta d$ are used for constructing keyword time series; Compared to $T_t$ and $T_d$ in \textbf{LED}, they do not have to adapt to the ``true'' scales of various events, thanks to the scale relationship model and the scale adjustment afterwards using the wavelet-based scheme. In practice, we can simply choose them to be relatively small, for example, as the expected minimum temporal and spatial intervals a desired event may span (specific example choices are presented in Sect.~\ref{sec:realresults}). Second, the number of spatial scales $n_\text{scale}$ can be considered as a choice in the design of the algorithm. Intuitively, an $n_\text{scale}$ too small would not take full advantage of the spatiotemporal scale relationship model, while $n_\text{scale}$ being too large might lead to unnecessary increase in computational cost. The choice of this parameter is also influenced by the resolution parameters $\Delta t$ and $\Delta d$. One the one hand, $\Delta d$ determines the number of geographical cells $l_d$ along one dimension hence the spatial variability in the data. This implicitly controls the maximum $n_\text{scale}$ such that the resulting distance scales are meaningful. On the other hand, given a certain time span of data, the temporal resolution $\Delta t$ would determine the length of keyword time series $l_t$, which in turn determines the maximum (meaningful) level of DWT computation using a Haar wavelet and hence the maximum temporal scale. Because of the relationship in Eq.~(\ref{eq:computescale}), the maximum spatial scale is thus determined accordingly. Based on these two observations, we therefore suggest considering $\lceil \text{min}(\text{log}_2{l_d},\text{log}_2{l_t}) \rceil$ as an upper bound for $n_\text{scale}$, where $\lceil \cdot \rceil$ denotes the ceiling of a number. In our experiments, we choose $n_\text{scale}=4$ to ensure a certain level of spatial variability while respecting this upper bound.


\section{Spatiotemporal analysis of noise in Twitter}
\label{sec:spatiotemporalnoise}
One challenge in designing event detection algorithms for Twitter data is that we often need to deal with a large amount of ``noise'' tweets that do not provide any information regarding real world events. Examples can be tweets such as ``Could really use a drink'' or ``Nachos for lunch'', or discussions between Twitter users about personal matters. We consider these tweets as {\it noise} and event detection algorithms should be able to discard them and not allow them to influence the event detection result. In the literature, several works (such as \citet{Sakaki10}) have employed keyword filtering techniques in order to tackle this problem and derived a working set of tweets that contain information relevant to the types of events they wish to detect. Since we do not focus in this paper on specific event types, but rather on events that take place in specific locations and time intervals, we analyze in this section the spatiotemporal structure of the noise, namely, the event-irrelevant tweets in the data. This analysis will allow us to define a novel term-filtering procedure, and to evaluate empirically the performance of the event detection algorithms in this paper using simulated noisy data under different space-time parameters.

\subsection{Spatial distribution of noise in Twitter data}
\label{sec:spatialanalysis}
\begin{center}
\begin{figure*}[t]
\subfigure[``nyc'']{
          \includegraphics[width=.49\textwidth]{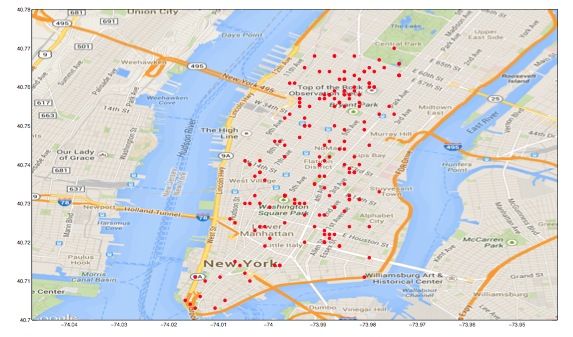}}
\subfigure[``love'']{
          \includegraphics[width=.49\textwidth]{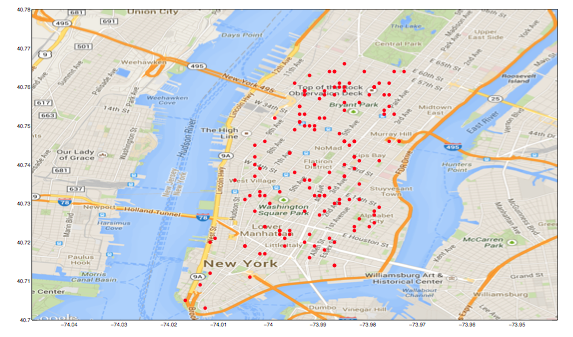}}
\subfigure[``lol'']{
          \includegraphics[width=.49\textwidth]{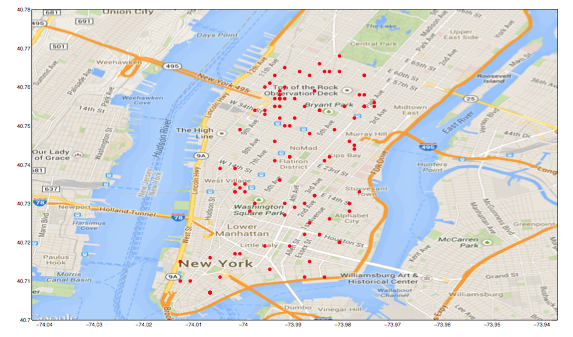}}
\subfigure[``night'']{
          \includegraphics[width=.49\textwidth]{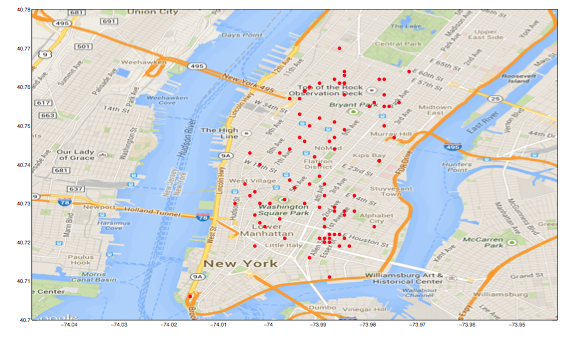}}
          \vspace{-0.2cm}
\caption[Locations of tweets that contain four specific frequent terms.]{Locations of tweets that contain four specific frequent terms.}	\label{fig:topTerms}
\end{figure*}
\end{center}

In order to get an intuition about the relevant spatial statistics models that can be useful for analyzing the spatial distribution of the noise, we focus on a set of geo-located tweets collected from a specific day (22-01-2012) in New York City. In this dataset, four of the top-ten frequent terms are: {\it nyc} contained in 335 tweets (183 of which are located in middle and lower Manhattan), {\it love} contained in 674 tweets (145 of which are located in middle and lower Manhattan), {\it lol} contained in 1080 tweets (110 of which are located in middle and lower Manhattan), and {\it night} contained in 355 tweets (97 of which are located in middle and lower Manhattan). These terms, albeit being among the most frequent ones in the daily collection of tweets, do not appear to be relevant to a specific event of interest. In Fig.~\ref{fig:topTerms} we illustrate the locations of the tweets (in middle and lower Manhattan) that contain these frequent terms. One can observe that the tweets have a slight, but not strong spatial concentration and appear to be almost randomly distributed within the Manhattan area. Based on these spatial plots we seek the appropriate spatial statistics tools to model these distributions. 

In the spatial statistics literature \citep{cressie+w_11}, the lack of spatial structure is commonly assessed using the concept of {\it Complete Spatial Randomness} (CSR). CSR considers that the points on a map (locations of tweets in our context) follow a homogeneous Poisson point process. This implies that the numbers of tweets in non-overlapping areas in the map are independent and follow a Poisson distribution with some intensity parameter $\lambda$. More precisely, if we denote the number of tweets within an area $A$ as $N(A)$, CSR asserts that $N(A)$ follows a Poisson distribution with mean $\lambda \cdot V(A)$, where $V(A)$ denotes the size of the area $A$. Intuitively, the CSR property asserts that points are ``randomly'' scattered in an area and are not concentrated in specific locations.

We consider the task of assessing the levels of noise in Twitter data (with respect to the target event detection task) by testing the CSR property for tweets that contain common terms\footnote{The direct usage of the CSR tests for the whole input tweet stream would not be particularly informative since both of our algorithms construct a similarity graph between tweets where the edge weights (i.e., the similarities between tweets) are based on the terms that two tweets have in common. In this case, noise or event-irrelevant tweets would affect the construction of the graph only when two ``noise'' tweets have a term in common (i.e., resulting in the formation of an edge that connects event-irrelevant tweets in the tweet similarity graph).}. In particular, we initially select a term (say the most frequent term in a collection of tweets) and then we test whether the locations of the tweets that contain this term have the CSR property. In case a term has the CSR property (i.e., the locations of the tweets that contain this term follow a Poisson point process distribution), the edges in the twitter similarity graph that are based on these terms can be considered as noise and may result in the identification of clusters that are not related to events of interest.


In order to evaluate the CSR property we have employed Ripley's $K$-function \citep{cressie+w_11}, which is a commonly used measure for assessing the proximity of a spatial distribution to a homogeneous Poisson point process. The sample-based estimate of Ripley's $K$-function is defined as $\widehat{K}(s)=V(A)\sum\limits_{i\neq j}N(d_{ij}<s)/n^2$ for a given distance value $s$, where $d_{ij}$ denotes the Euclidean distance between two sample points $i$ and $j$ (two tweets in our context) in the space, $N(d_{ij}<s)$ counts the number of sample pairs that has a distance smaller than $s$, $n$ is the total number of points, and $V(A)$ is the size of the area $A$. It is known that, when a spatial Poisson process is homogeneous, the values of the $K$-function are approximately equal to $\pi s^2$. Thus, the proximity of $\widehat{K}(s)$ to $\pi s^2$ can be employed for evaluating how similar our data distribution is to a homogeneous Poisson process. In this paper, we use the standardized $K$-function: $\widehat{L}(s)=\sqrt{\frac{\widehat{K}(s)}{\pi}}-s$, and the proximity to a homogeneous Poisson process is measured by the proximity of the values of $\widehat{L}(s)$ to 0.

\begin{center}
\begin{figure*}[t]
\subfigure[``nyc'']{
          \includegraphics[width=.49\textwidth]{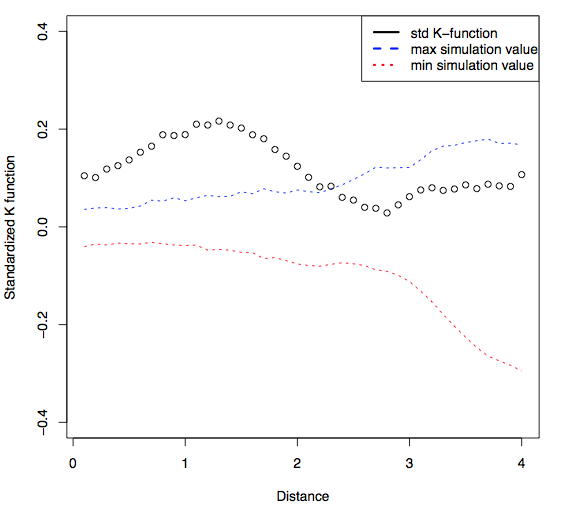}}
\subfigure[``love'']{
          \includegraphics[width=.49\textwidth]{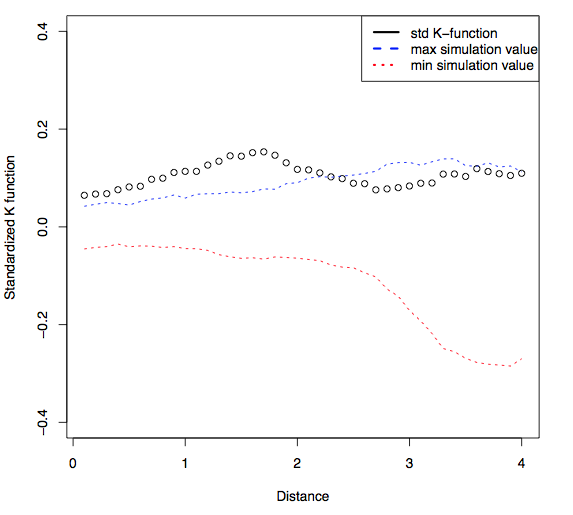}}
\subfigure[``lol'']{
          \includegraphics[width=.49\textwidth]{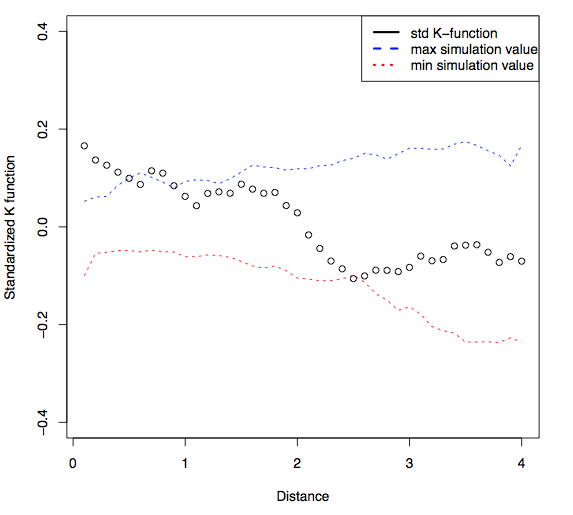}}
\subfigure[``night'']{
          \includegraphics[width=.49\textwidth]{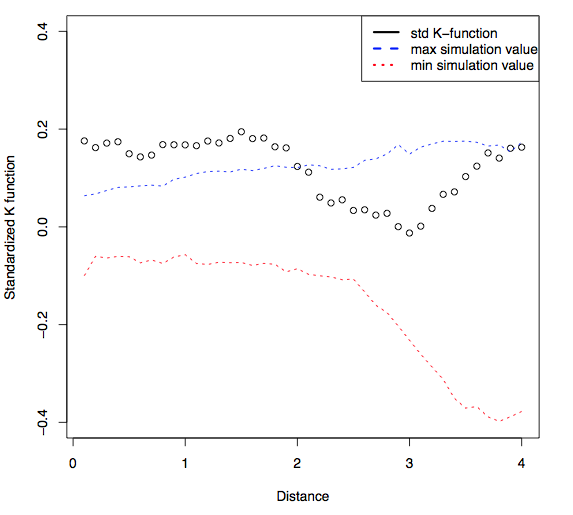}}
          \vspace{-0.2cm}
\caption[Comparison between the sample-based estimates of the standardized $K$-function for tweets containing the four specific terms (black lines), and the max-min values of this function for simulated homogeneous Poisson processes (red and blue lines).]{Comparison between the sample-based estimates of the standardized $K$-function for tweets containing the four specific terms, and the max-min values of this function for simulated homogeneous Poisson processes.}	\label{fig:Kfunction}
\end{figure*}
\end{center}

We now assess the spatial distribution of the sets of tweets shown in Fig.~\ref{fig:topTerms} (tweets containing the terms ``nyc'', ``love'', ``lol'' and ``night''). Specifically, we illustrate in Fig.~\ref{fig:Kfunction} the values of their standardized $K$-function for different values of $s$ (distances) up to 4km, depicted in the black lines. Moreover, we simulate (2000 times) a homogeneous Poisson process and compute the maximum and minimum values for $\widehat{L}(s)$, depicted in the blue and red dashed lines, respectively. We can observe that, the values of $\widehat{L}(s)$ obtained using the locations of these tweets are close to, and in several cases within the ranges of, the values of $\widehat{L}(s)$ obtained from the simulated homogeneous Poisson processes. This indicates that these tweets are slightly more concentrated in space than what a homogeneous Poisson process would produce (possibly due to the differences in the concentration of twitter users in different areas in middle and lower Manhattan), but their spatial distribution is still close to a homogeneous Poisson process.

To further explain what we mean by ``still close to a homogeneous Poisson process'', let us consider what appears to be one of the most extreme differences between the spatial distribution of tweets and a homogeneous Poisson process in Fig.~\ref{fig:Kfunction}, which is the value $\widehat{L}(s)=0.19$ that is achieved for a distance value $s=1km$ for the term ``nyc''. Based on the number of tweets that contain the term ``nyc'' on 22-01-2012 (in middle and lower Manhattan), a homogeneous Poisson process would require an intensity parameter $\lambda=7.93$ per square kilometer to generate the same number of tweets. This would mean that on average, the number of tweets per square kilometer that contain the term ``nyc'' should be $7.93$. In our case, the value of $\widehat{L}(s)=0.19$ for $s=1km$ means that, for small distances, the actual concentration of tweets is slightly higher, with an intensity parameter $\lambda=11.21$ per square kilometer. This shows that, even in this worst case, the spatial distribution of tweets is still not far from a homogeneous Poisson process.

\begin{figure}[t]
	\begin{center}
		\begin{tabular}{cc}
			\includegraphics[width=.80\textwidth]{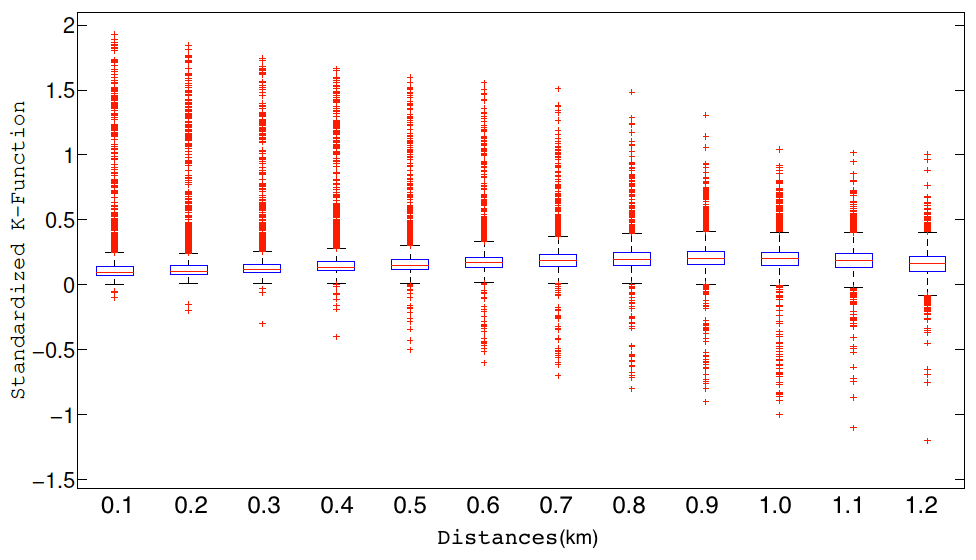}~\\
		\end{tabular}
	\end{center}
	\vspace{-0.4cm}
	\caption{Boxplot of the values of the standardized $K$-function for the most frequent terms.}
	\label{fig:BoxPlot}
\end{figure}

In order to evaluate whether our observation for the four specific terms holds for a larger tweet collection, we analyze all the geo-located tweets from the New York area for the duration between 01-11-2011 and 01-04-2013. Specifically, for each day, we have retrieved the top-ten frequent terms, and for each frequent term we have computed the sample-based estimates of $\widehat{L}(s)$ for $s$ from 0.1km to 1.2km, again focusing on the middle and lower Manhattan area. To avoid cases where the number of samples is low, we have computed the values of $\widehat{L}(s)$ only when the number of tweets in middle and lower Manhattan is larger than 100. The results are presented in the boxplot of Fig.~\ref{fig:BoxPlot}, which illustrates the mean, the variance and the range of the values of $\widehat{L}(s)$ (around 5000 values in total, ten for each of the \~500 days), for different values of $s$. As we can see, the boxplot in Fig.~\ref{fig:BoxPlot} illustrates that the most frequent terms in our Twitter data do not have a strong spatial pattern and follow a distribution that is close to a homogeneous Poisson process, only exhibiting slightly higher tweet concentrations for small distances.


%

\subsection{Temporal distribution of noise in Twitter data}
\label{sec:temporalanalysis}
In order to analyze the temporal pattern of the noise in Twitter data, we have assessed whether the distribution of the timestamps of event-irrelevant tweets is close to a uniform distribution. A uniform distribution of the timestamps can serve as a strong indication that these tweets are not relevant to an event that takes place in a confined time interval. In order to test this hypothesis, we have collected the timestamps of the top-ten frequent terms of each day between 01-11-2011 and 01-04-2013. We focus our analysis on a 6-hour interval between 11am and 5pm. For this time interval we tested whether the timestamps of tweets that contain a specific frequent term follow a uniform distribution, using the Chi-squared goodness of fit test. Interestingly, we could reject the null hypothesis that the timestamps are uniformly distributed at a 5\% confidence level only in 27\% of the cases. This result suggests that a large number of frequent terms in our data does not have a strong temporal pattern.

In summary, the spatiotemporal analysis of the distribution of the noise in Twitter data presented in this section allows us to (i) conduct synthetic experiments with simulated noisy data that help us understand the behavior of the event detection algorithms under different space-time parameters, and (ii) consider a term-filtering mechanism that removes tweets that contain the terms with low values for $\widehat{L}(s)$. We will describe both aspects in more details in the next section.

\section{Synthetic experiments}
\label{sec:synresults}
In order to better understand the behavior of the event detection algorithms \textbf{LED} and \textbf{MED}, and the potential influence of the noise in the data, we present in this section experimental results based on synthetic data. Specifically, we generate artificial documents that are considered as ``tweets'' posted at different time instants and diverse spatial locations. By creating some artificial ``events'' in this setting, we are able to evaluate quantitatively the performances of the proposed methods under different choices for the parameter values. In what follows, we first explain the experimental setup, and then present the event detection results.

\subsection{Experimental setup}
We work with a spatial area of 10 by 10, which are defined by bottom left and top right coordinates (0, 0) and (10, 10) respectively in a 2-D Euclidean space, and a temporal interval of (0, 32) on the real line. We then define events that span different spatial areas and temporal intervals in diverse experimental settings. First, for each event, we choose a number between 3 and 10 uniformly at random as the number of tweets related to that event. These event-relevant tweets are uniformly distributed in the spatial area and temporal interval spanned by that event. We also generate, based on the spatiotemporal analysis presented in Sect.~\ref{sec:spatiotemporalnoise}, event-irrelevant tweets, namely, noise, which follows a 2-D Poisson point process in the whole spatial area and are distributed uniformly in the whole temporal interval. Next, the content of each tweet is generated as follows. We take geo-located tweets from New York collected on a random day (in this case 21-01-2012) as a reference, and choose 59 terms as event-relevant terms (referred to as signal terms) and consider all the other terms that appear in the tweets on that day as noise (referred to as noise terms). We select the number of terms in each event-relevant tweet uniformly at random between 5 and 10. In particular, in each event-relevant tweet, one term is selected uniformly at random from the 59 signal terms, and the rest are randomly chosen from the noise terms with probabilities that depend on their numbers of occurrences in the actual daily tweets. We also create event-irrelevant tweets, and the number of terms in each event-irrelevant tweet is selected uniformly at random between 3 and 10. The terms in each event-irrelevant tweet are only chosen from the noise terms. We present event detection results in the following scenarios.

\subsection{Event detection results in synthetic data}
\subsubsection{Events concentrated in both time and space without noise}
\label{sec:synresults1}
\begin{figure}[t]
	\begin{center}
		\begin{tabular}{cc}
			~\includegraphics[width=0.45\textwidth]{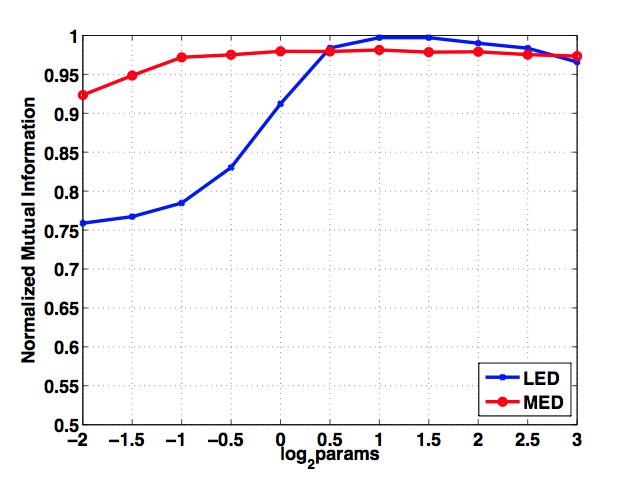}~ & ~\includegraphics[width=0.45\textwidth]{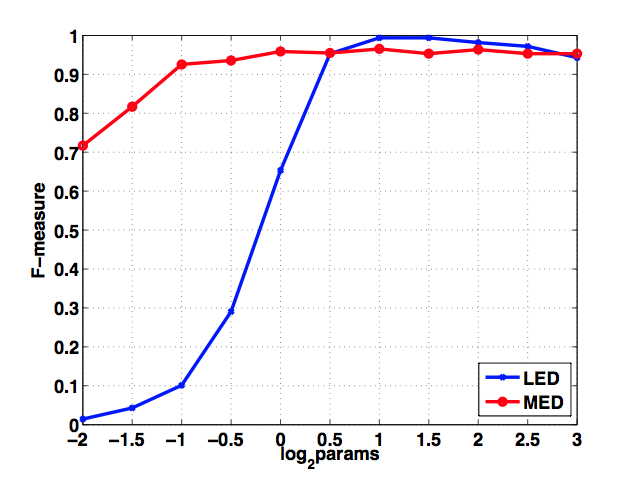}~\\
			~(a)~ & ~(b)~\\
		\end{tabular}
	\end{center}
	\vspace{-0.4cm}
	\caption{Clustering performance in terms of (a) NMI and (b) F-measure, on events concentrated in both time and space without noise.}
	\label{fig:synresults1}
\end{figure}

In a first scenario, we consider 20 events, each of which is concentrated in a 2 by 2 spatial area and a temporal interval of 2. The spatial and temporal locations are chosen uniformly at random in the whole spatial area and temporal interval. We only consider event-relevant tweets, where the goal is to detect the 20 clusters that correspond to the events by clustering the tweets into different subsets. For \textbf{MED}, we focus on terms that appear in at least 3 tweets. We choose $n_\text{scale}=4$ unless its upper bound $\lceil \text{min}(\text{log}_2{l_d},\text{log}_2{l_t}) \rceil$ goes below 4 due to the increase of resolution parameters. In our experiments, we take the same value for the four parameters in the two methods, namely $T_t$ and $T_d$ in \textbf{LED} and $\Delta t$ and $\Delta d$ in \textbf{MED}, and evaluate the clustering performance in terms of \textit{Normalized Mutual Information (NMI)} and \textit{F-measure} \citep{Manning08}. The \textit{F-measure} is computed using a choice of $\beta=2$ meaning that it is slightly in favor of recall\footnote{\textit{F-measure} is computed as $(1+\beta^2) \cdot \frac{\textit{Precision} \cdot \textit{Recall}}{(\beta^2 \cdot \textit{Precision}) + \textit{Recall}}.$}, as we consider that it is more important to ensure that tweets related to the same event are grouped into one cluster. The results obtained by averaging 10 test runs are shown in Fig.~\ref{fig:synresults1}. As we can see, in terms of both evaluation criteria, the performance of \textbf{LED} with small values of the thresholds $T_t$ and $T_d$ is not satisfactory as it is not able to capture the links between all the tweets within the same event. However, the performance increases noticeably as the temporal and spatial thresholds are chosen to be close to or larger than the ``true'' scales of the events (2 in this case for both time and space). When the thresholds get too large, the performance drops slightly, as the chance of grouping two different events together in one cluster increases. Compared to \textbf{LED}, \textbf{MED} achieves much better performance even when the resolution parameters $\Delta t$ and $\Delta d$ are small. The reason is that, even at very fine initial resolutions, the wavelet-based representation in \textbf{MED} is able to aggregate the time series appropriately such that the similarity of the time series is actually computed at a coarser scale. This suggests that \textbf{MED} is better at capturing the links between tweets corresponding to the same event, even with suboptimal choices for the value of the parameters. Therefore, the performance of \textbf{MED} is much less sensitive to parameter selection than that of \textbf{LED}.

\subsubsection{Events concentrated in only one dimension without noise}
\label{sec:synresults2}
\begin{figure}[t]
	\begin{center}
		\begin{tabular}{cc}
			~\includegraphics[width=0.45\textwidth]{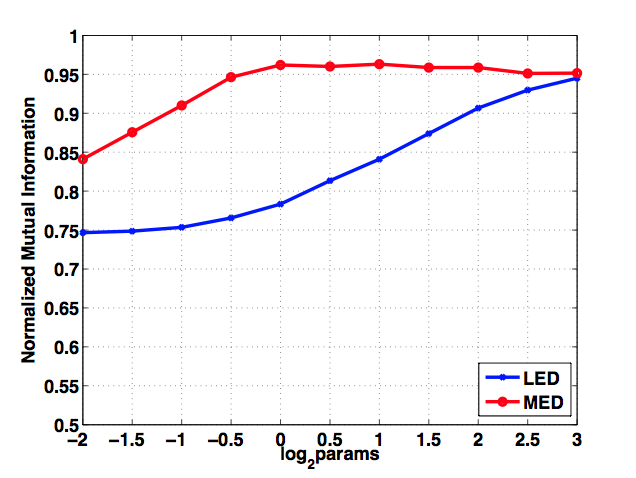}~ & ~\includegraphics[width=0.45\textwidth]{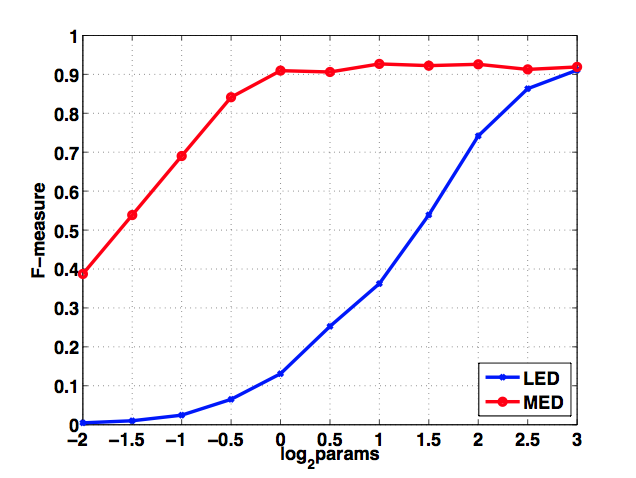}~\\
			~(a)~ & ~(b)~\\
		\end{tabular}
	\end{center}
	\vspace{-0.4cm}
	\caption{Clustering performance in terms of (a) NMI and (b) F-measure, on events concentrated in only one dimension without noise.}
	\label{fig:synresults2}
\end{figure}

We now consider events that are not necessarily concentrated in both time and space but only in one of the two dimensions. Specifically, we consider 20 events, where 10 of them are concentrated in a temporal interval of length between 1 to 2 but spread in a spatial area with a size from 8 by 8 to 16 by 16. The other 10 events are concentrated in a spatial area with a size from 1 by 1 to 2 by 2 but spread in a temporal interval of length between 8 to 16. We still consider a noise-free scenario as in the previous experiment. The clustering results are shown in Fig.~\ref{fig:synresults2}. We see that, while \textbf{MED} can handle the scale changes in this scenario with a performance that remains comparable to that in the previous experiment, the performance of \textbf{LED} drops significantly. Specifically, due to the lack of a single temporal and spatial scale for all the events, \textbf{LED} only performs reasonably well when the threshold values for $T_t$ and $T_d$ are large enough to cover the scales of all the events. This experiment highlights the advantage of \textbf{MED} in handling events of different scales and in the absence of simultaneous temporal and spatial localization.

\subsubsection{Events concentrated in both time and space with noise}
\label{sec:synresults3}
We now move to noisy scenarios where we also consider event-irrelevant tweets in addition to event-relevant tweets. Specifically, we generate event-irrelevant tweets that follow a 2-D Poisson point process with an intensity parameter $\lambda=10$ within the whole spatial area of 10 by 10. This generates around 1000 noise tweets in addition to the tweets that correspond to 20 events generated as in Sect.~\ref{sec:synresults1}. The goal is to detect the events by applying clustering to all the tweets in the dataset. To measure the clustering quality, we define the groundtruth to be a combination of 20 event clusters and noise clusters where each noise tweet is considered as a single cluster. The reason for this setting is that we wish to group tweets that correspond to the same event, and at the same time we want to ensure that the noise tweets remain as separated as possible. Based on the analysis in Sect.~\ref{sec:spatiotemporalnoise}, for \textbf{MED}, we propose to evaluate the values of the standardized $K$-function $\widehat{L}(s)$ for all the terms that appear in at least 3 tweets for $s$ chosen to be 0.5, 1, 1.5 and 2, and only consider terms that have an average $\widehat{L}(s)$ value no smaller than 1 as valid terms for generating keyword time series.  
The clustering results are shown in Fig.~\ref{fig:synresults3}. In the noisy scenario, we see that the \textit{NMI} and \textit{F-measure} curves show different trends. Specifically, with small values for the threshold or resolution parameters, the number of links between tweets created by both methods is small, and most of the noise clusters remain well-separated. When the parameter values increase, noise tweets starting forming more links to event-relevant tweets as well as to themselves, which penalizes the clustering.
Therefore, we see that the \textit{NMI} curves show an almost monotonically decreasing trend as the parameter values increase. In contrast, the \textit{F-measure} is a weighted combination of precision and recall, which penalizes both false positives and false negatives. Therefore, for both methods, we see that the \textit{F-measure} curves initially increase as the parameter values increase (where the number of false negatives generally decreases), and decrease as these parameters become large (where the number of false positives increases).

We now compare the performance of \textbf{LED} and \textbf{MED} in the same experiment. For \textit{NMI}, we see that the performance of \textbf{LED} drops significantly when the thresholds exceed the ``true'' scales of the events, as large thresholds in \textbf{LED} tend to increase the number of event-relevant and noise tweets that are linked to each others. In comparison, the performance of \textbf{MED} is relatively more stable, which is partly due to the term-filtering procedure employed. Similarly, we see that \textbf{MED} outperforms \textbf{LED} for a large range of parameter values in terms of the \textit{F-measure}. In addition, the performance of \textbf{MED} is again more stable in the sense that it peaks at a wider range of parameter values, while \textbf{LED} only performs well when the threshold values are chosen at the ``true'' event scales.


\subsubsection{Events concentrated in only one dimension with noise}
\label{sec:synresults4}
\begin{figure}[t]
	\begin{center}
		\begin{tabular}{cc}
			~\includegraphics[width=0.45\textwidth]{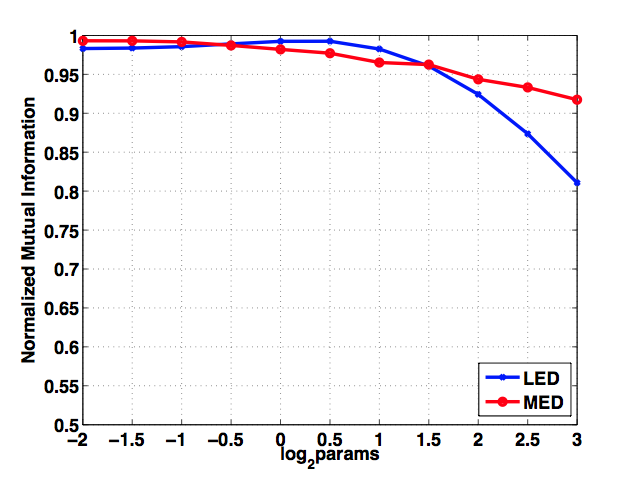}~ & ~\includegraphics[width=0.45\textwidth]{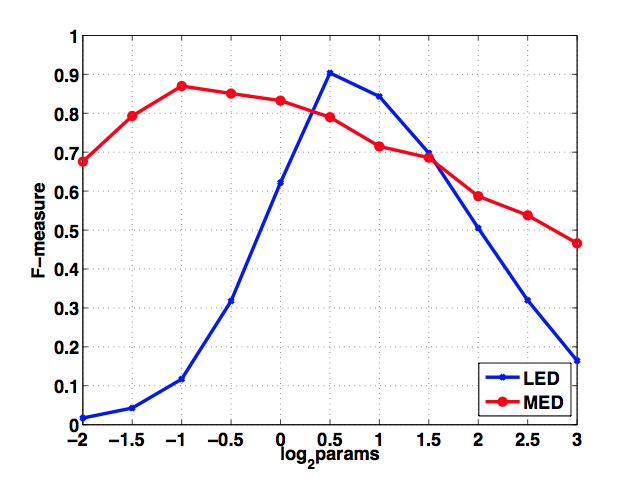}~\\
			~(a)~ & ~(b)~\\
		\end{tabular}
	\end{center}
	\vspace{-0.4cm}
	\caption{Clustering performance in terms of (a) NMI and (b) F-measure, on events concentrated in both time and space with noise.}
	\label{fig:synresults3}
\end{figure}

\begin{figure}[t]
	\begin{center}
		\begin{tabular}{cc}
			~\includegraphics[width=0.45\textwidth]{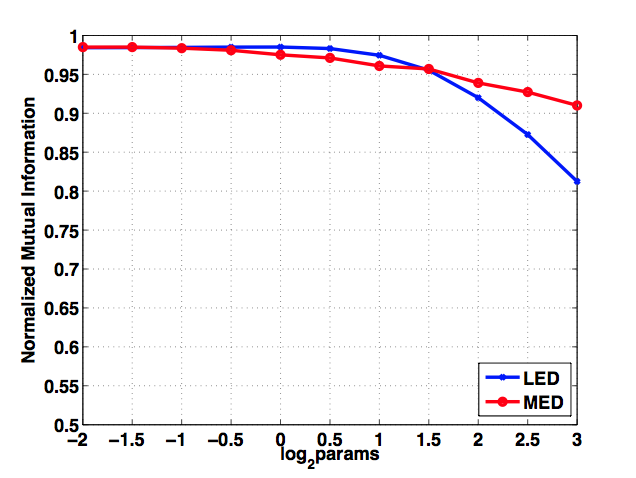}~ & ~\includegraphics[width=0.45\textwidth]{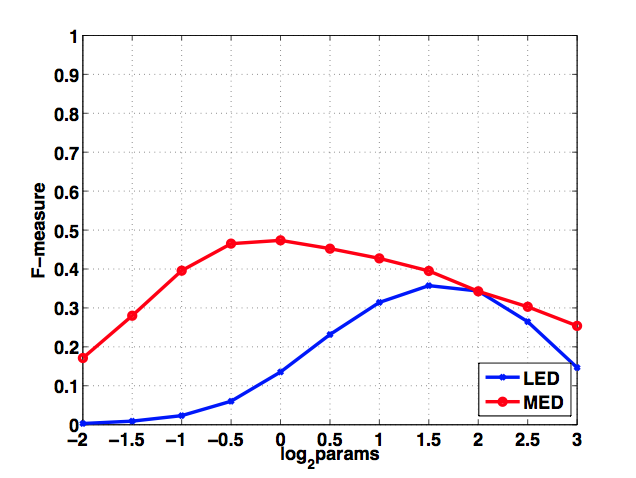}~\\
			~(a)~ & ~(b)~\\
		\end{tabular}
	\end{center}
	\vspace{-0.4cm}
	\caption{Clustering performance in terms of (a) NMI and (b) F-measure, on events concentrated in only one dimension with noise.}
	\label{fig:synresults4}
\end{figure}

Finally, we show in Fig.~\ref{fig:synresults4} the experimental results in a noisy scenario where the events are concentrated either in time or space as defined in Sect.~\ref{sec:synresults2}. While the \textit{NMI} curves are similar to those in Fig.~\ref{fig:synresults3}, the \textit{F-measure} curves show that the performance of both methods drops significantly in this challenging scenario. Still, \textbf{MED} outperforms \textbf{LED} in terms of both peak performance and stability.

\subsubsection{Influence of parameter settings}
We now take a closer look at the parameter settings for the synthetic experiments. Especially, we investigate how the length of the temporal interval, the size of the spatial area, and the number of signal terms in each event-relevant tweet, influence the performance of both algorithms in terms of the \textit{F-measure} in the scenario of Sect. 6.2.4, that is, the performance curve in Fig.~\ref{fig:synresults4}(b)).

\begin{figure}[t]
      \centering
            \subfigure[]	{ \includegraphics[width=0.30\textwidth]{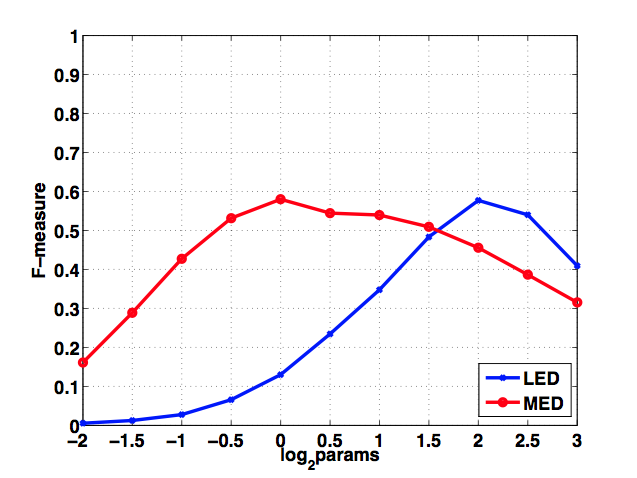}   	\label{}}
            \subfigure[] 	{ \includegraphics[width=0.30\textwidth]{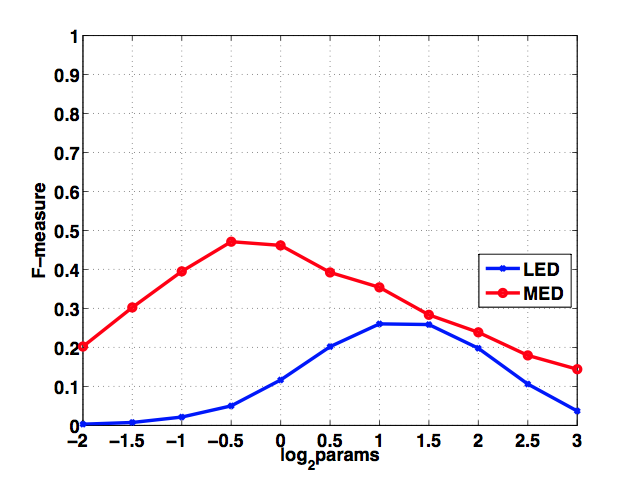}	\label{}}
            \subfigure[] 	{ \includegraphics[width=0.30\textwidth]{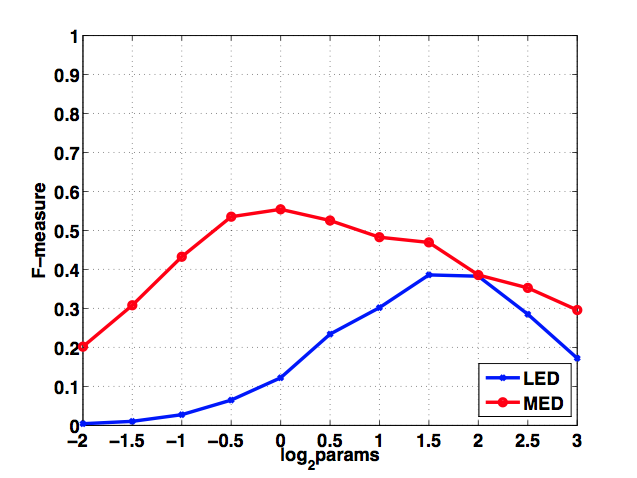}	\label{}}
        \vspace{-0.2cm}
        \caption{Clustering performance in terms of F-measure on events concentrated in only one dimension with noise. (a) The same setting as Sect. 6.2.4. but with a temporal interval of 128. (b) The same setting as Sect. 6.2.4. but with a spatial area of 16 by 16. (c) The same setting as Sect. 6.2.4. but with 3 signal terms in each event-relevant tweet.}
        \label{fig:params}
\end{figure}

First, given a fixed parameter for the Poisson point process and fixed spatial area of 10 by 10, the total number of noise tweets remains the same. In this case, we observe that the performance of both algorithms has improved when the temporal interval increases from 32 to 128 (Fig.~\ref{fig:params}(a)), due to decreased noise density in the temporal dimension hence a higher signal-to-noise-ratio. Such a gain is more dramatic for \textbf{LED} especially at large parameter values, in which case the performance of this approach is more sensitive to the density of the noisy information.

Second, given a fixed temporal interval of 32, as the spatial area increases from 10 by 10 to 16 by 16, the total number of noise tweets increases quadratically. 
In this case, we see from Fig.~\ref{fig:params}(b) that the performance of both algorithms decreases mainly because of that, as the total number of noise tweets increases, generally more links are formed between noise tweets.

Finally, we have investigated the influence of the number of signal terms in each event-relevant tweet on the performance of the algorithms. Specifically, we increase the number of signal terms from 1 to 3 in each event-relevant tweet and repeat the same experiments. We have observed performance gain in Fig.~\ref{fig:params}(c) for both algorithms which matches the intuition that a higher signal-to-noise-ratio generally leads to better performance.

In summary, the synthetic experiments suggest that \textbf{LED} is efficient at detecting events that are concentrated in both time and space, provided that these events are of similar scales and that the correct temporal and spatial thresholds are chosen in the algorithm. In comparison, although we employed a term-filtering procedure in \textbf{MED} in the noisy scenarios, the results on synthetic data generally suggest that \textbf{MED} is better than \textbf{LED} at detecting events of different scales and in the absence of simultaneous temporal and spatial localization. \textbf{MED} is also less sensitive to parameter selection and leads to more robust and stable event detection performance.

\section{Real world experiments}
\label{sec:realresults}
We now test the performance of \textbf{LED} and \textbf{MED} in real world event detection tasks. We focus in this section on the comparison between these two event detection methods, since (i) such a comparison would highlight the difference between \textbf{LED} and \textbf{MED} in detecting real world events of various temporal and spatial scales, and (ii) to the best of our knowledge, there is no other multiscale method in the literature that is dedicated to event detection. We first describe the data and some implementation details, and then present the event detection results. Finally, we discuss about the scalability of the proposed algorithm.

\subsection{Data description}
We have collected geotagged public tweets in the New York area, which corresponds to a geographical bounding box with bottom left GPS coordinates pair (40.4957, -74.2557) and top right coordinates pair (40.9176, -73.6895), from November 2011 to March 2013. The streams of public tweets are retrieved using Twitter's official Streaming API with the ``locations'' request parameter\footnote{\url{https://dev.twitter.com/streaming/overview/request-parameters#locations}}. After the initial retrieval, we filter out those tweets that have no geotags or have geotags outside the predefined bounding box. This results in 16449769 geotagged tweets in total. As a pre-processing step, we remove those tweets that contain a clear location indicator, such as the ones corresponding to Foursquare check-ins, which we do not consider as events of interest.

\subsection{Implementation details}
We implement both event detection algorithms \textbf{LED} and \textbf{MED} on a daily basis, that is, we aim at detecting events from each day.  The \textit{tf-idf} weighting scheme in the vector space model is implemented using the Text to Matrix Generator (TMG) MATLAB toolbox \citep{Zeimpekis06}, where we also remove a list of stop words provided by the toolbox (with an additional one ``http''), and set the minimum and maximum length of a valid term to be 3 and 30.

For \textbf{LED}, we use a temporal threshold of $T_t=30$ minutes and spatial threshold of $T_d \approx 100$ meters (difference of 0.001 in latitude or 0.0015 in longitude) in Eq.~(\ref{eq:local}) for the detection of local event clusters. For \textbf{MED}, we focus on terms that appear in at least 5 tweets. We evaluate the values of the standardized $K$-function $\widehat{L}(s)$ for all these terms with $s$ chosen to be 0.2, 0.4, 0.6, 0.8 and 1, and only consider those that have an average $\widehat{L}(s)$ value no smaller than 0.5  as valid terms for generating keyword time series. The initial temporal and spatial resolutions in \textbf{MED} are set to $\Delta t=30$ minutes and $\Delta d=100$ meters, and the number of spatial scales is set to $n_\text{scales}=4$. Once the clusters are obtained by both methods, we perform simple post-processing steps that (i) remove clusters that contain less than 3 tweets or less than 3 distinct users, so that each event would contain sufficient information from sufficient number of observes, and (ii) remove clusters in which more than 50\% of the tweets comes from a single user, so that the information source is sufficiently diverse, and finally (iii) remove clusters that correspond to job advertisements and traffic alerts posted by bots. While there is no general rule for such post-processing, we found these steps practical to remove clusters that are not meaningful and correspond to noisy information.

\begin{table}[t]
	\caption{Example local clusters detected by \textbf{LED} that correspond to protests in the OWS movement.}
	\begin{center}
		\begin{tabular}{cc}
			~\includegraphics[width=0.98\textwidth]{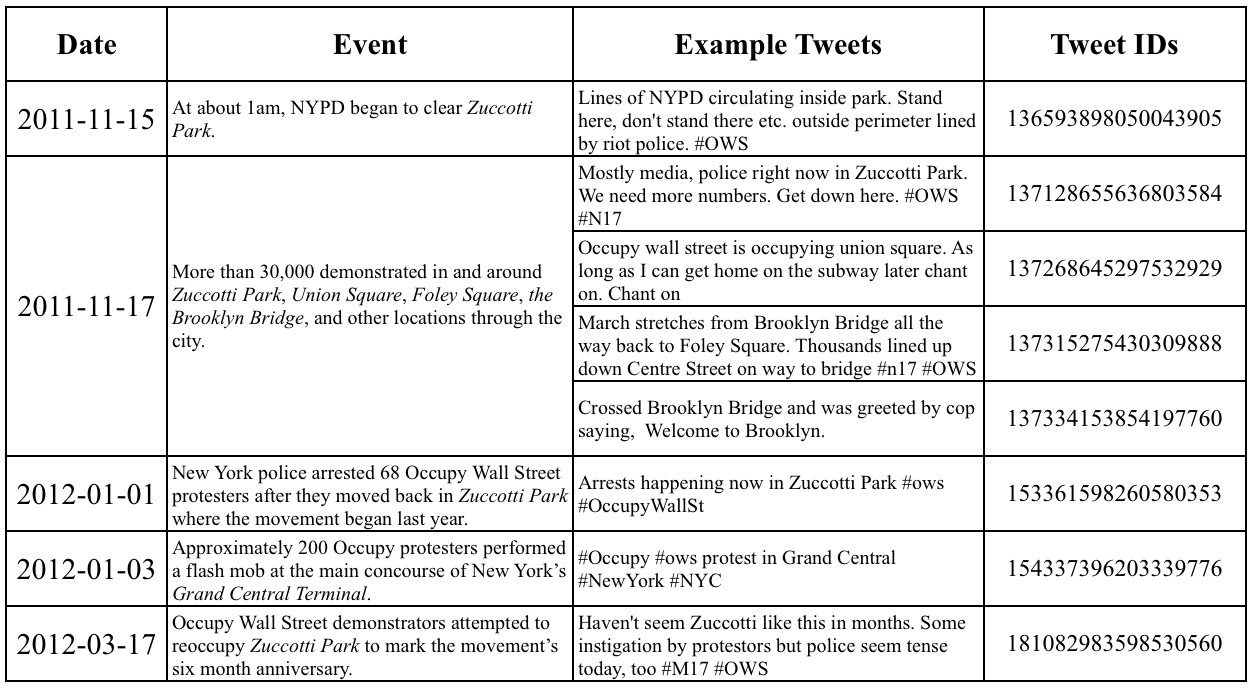}~\\
		\end{tabular}
	\end{center}
	\label{tab:owsevents}
\end{table}

\subsection{Event detection results}
\begin{figure}[t]
	\begin{center}
		\begin{tabular}{cc}
			~\includegraphics[width=0.95\textwidth]{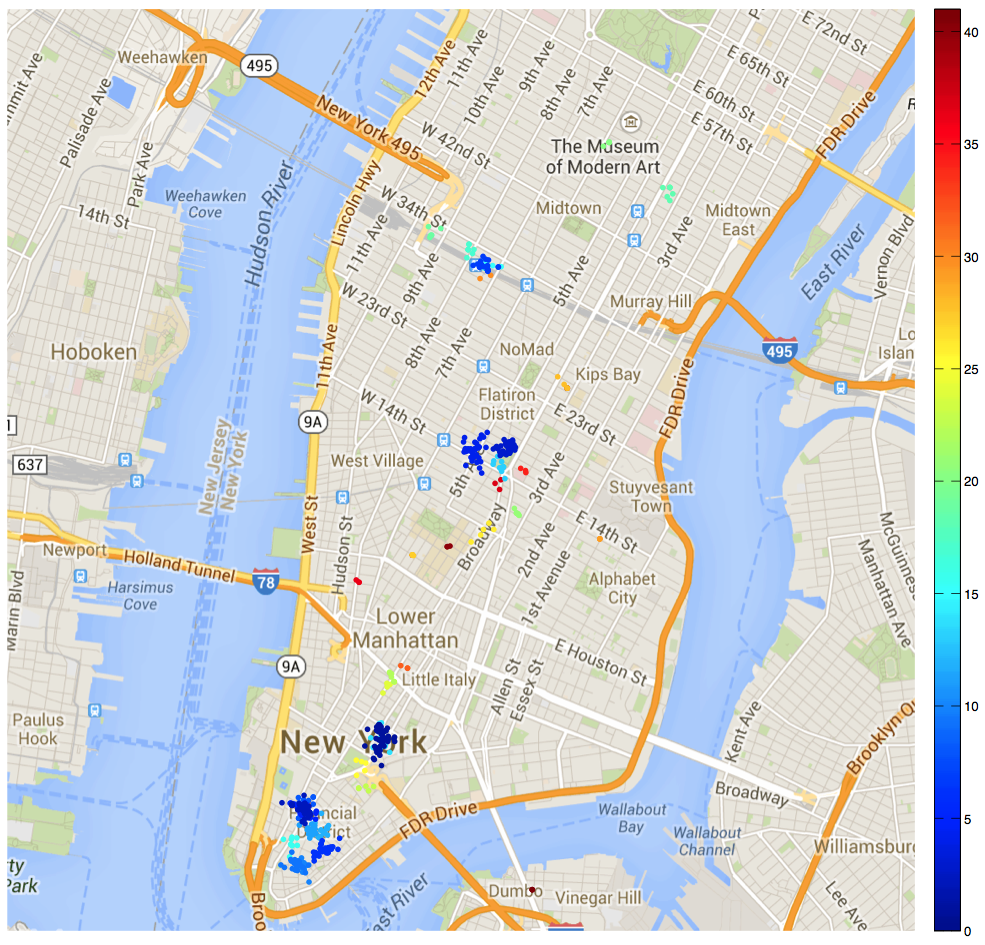}~\\
		\end{tabular}
	\end{center}
	\vspace{-0.4cm}
	\caption{Local event clusters detected by \textbf{LED} on 2011-11-17. The colors represent cluster ids.}
	\label{fig:st_clusters}
\end{figure}

We now analyze the clustering results for both \textbf{LED} and \textbf{MED} algorithms. First of all, the clusters detected by \textbf{LED} do correspond to meaningful real world events of interest. For example, Table~\ref{tab:owsevents} shows some example local clusters obtained that correspond to several protests during the Occupy Wall Street (OWS) movement\footnote{\url{http://en.wikipedia.org/wiki/Occupy_Wall_Street}} in New York City. To understand better the behavior of \textbf{LED}, we take 2011-11-17 as an example date, when many OWS protests took place. We first show in Fig.~\ref{fig:st_clusters} all the 41 local event clusters detected on this date in middle and lower Manhattan, where different clusters are shown in different colors. Detailed information about the top 20 clusters are further shown in Table~\ref{tab:st_clustersinfo}, where the six columns correspond to cluster id, median timestamp (GMT+0) of all the tweets in the cluster, minimal time interval (in seconds) that covers 80\% of the tweets, mean latitude and longitude of the tweets, and (up to 10) top terms contained in the cluster. As we can see in Fig.~\ref{fig:st_clusters} and in the third column of Table~\ref{tab:st_clustersinfo}, all the clusters are highly localized in both time and space. In addition, due to the strict temporal and spatial constraints used by \textbf{LED} (see Eq.~(\ref{eq:local})), for the same event we get separate clusters, which correspond to different timestamps (such as clusters 2 and 5 that talk about protests at Zuccotti Park) or different locations (such as clusters 3 and 13 that talk about protests at Union Square). Ideally, we would like some of these separated clusters to be grouped together if they are related to the same real world event.

\begin{table}[t]
	\caption{Detailed information about the top 20 local event clusters detected by \textbf{LED} on 2011-11-17. The six columns from the left to the right correspond to cluster id, median timestamp (GMT+0) of all the tweets in the cluster, minimal time interval (in seconds) that covers 80\% of the tweets, mean latitude and longitude of the tweets, and (up to 10) top terms contained in the cluster.}
	\begin{center}
		\begin{tabular}{cc}
			~\includegraphics[width=0.98\textwidth]{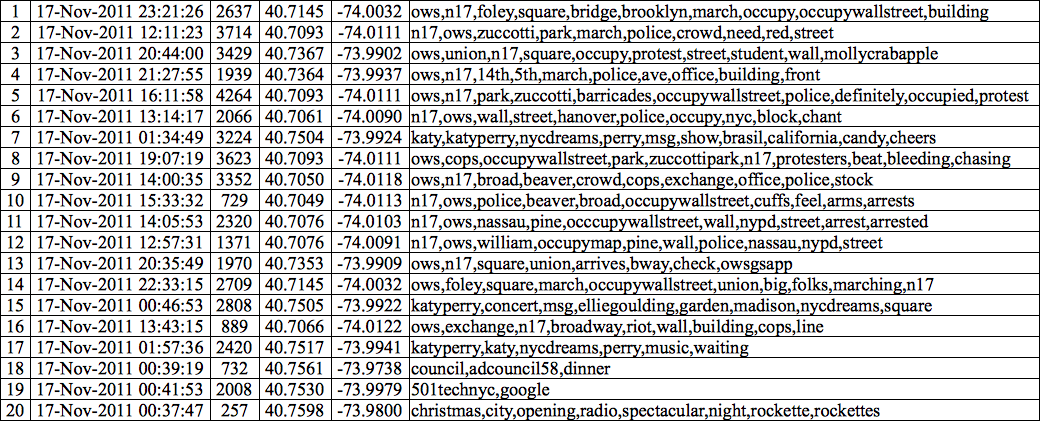}~\\
		\end{tabular}
	\end{center}
	\label{tab:st_clustersinfo}
\end{table}

\begin{table}[t]
	\caption{Detailed information about the top 10 event clusters detected by \textbf{MED} on 2011-11-17. The six columns from the left to the right correspond to cluster id, median timestamp (GMT+0) of all the tweets in the cluster, minimal time interval (in seconds) that covers 80\% of the tweets, mean latitude and longitude of the tweets, and (up to 10) top terms contained in the cluster.}
	\begin{center}
		\begin{tabular}{cc}
			~\includegraphics[width=0.98\textwidth]{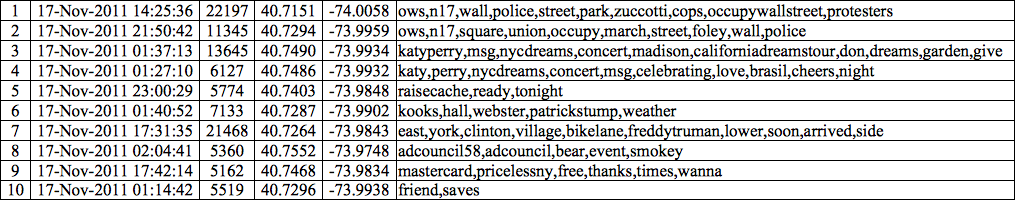}~\\
		\end{tabular}
	\end{center}
	\label{tab:wavelet_clustersinfo}
\end{table}

\begin{figure}[t]
	\begin{center}
		\begin{tabular}{cc}
			~\includegraphics[width=0.45\textwidth]{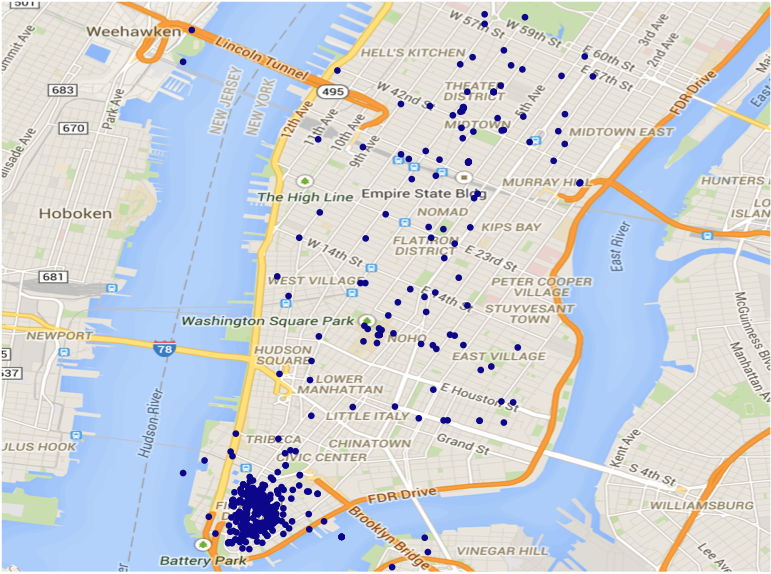}~ & ~\includegraphics[width=0.45\textwidth]{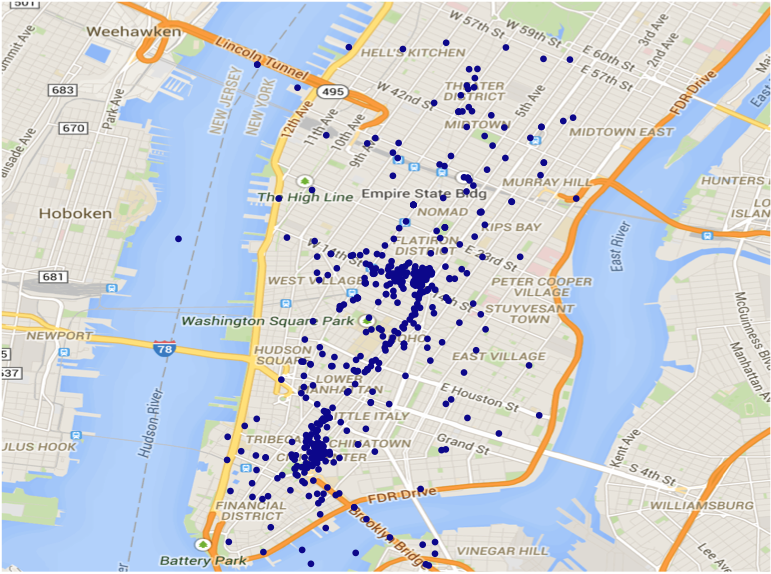}~\\
			~(a)~ & ~(b)~\\
			~\includegraphics[width=0.45\textwidth]{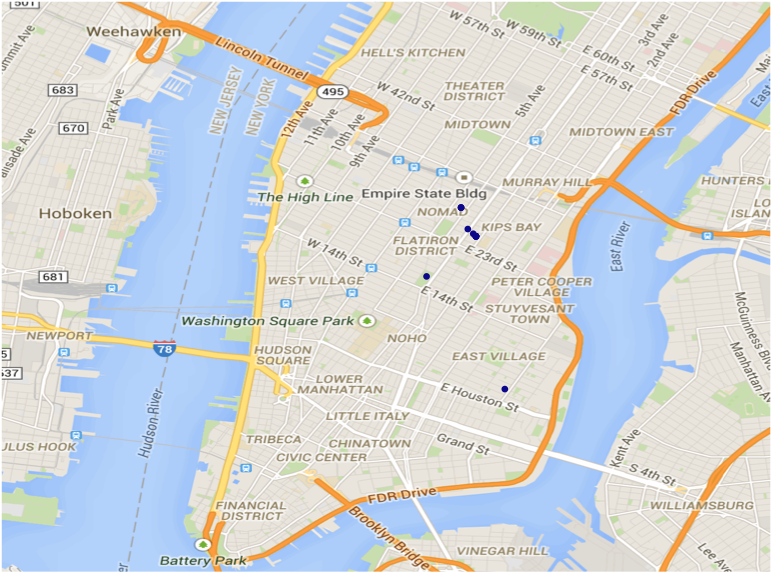}~ & ~\includegraphics[width=0.45\textwidth]{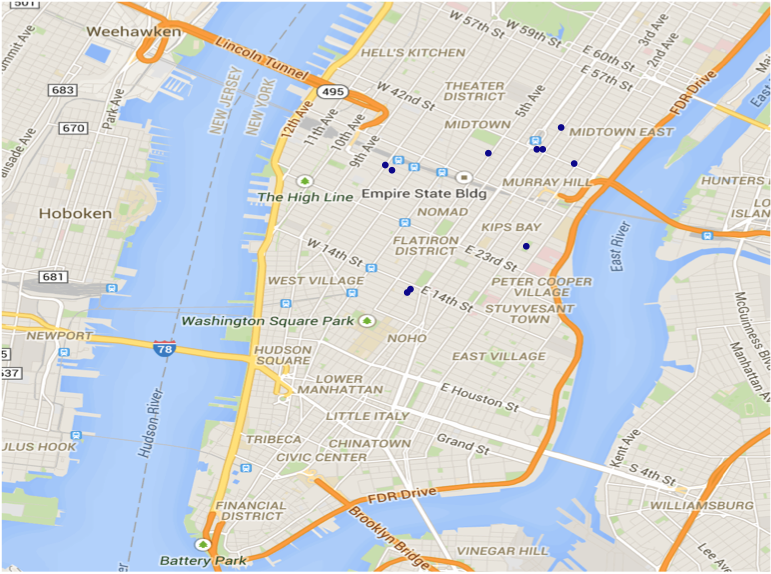}~\\
			~(c)~ & ~(d)~\\
		\end{tabular}
	\end{center}
	\vspace{-0.4cm}
	\caption{Example event clusters detected by \textbf{MED} on 2011-11-17: (a) OWS protests at Zuccotti Park (cluster 1). (b) OWS protests at Union Square and Foley Square (cluster 2). (c) Raise Cache (cluster 5). (d) Mastercard free lunch promotion (cluster 9).}
	\label{fig:wavelet_clusters}
\end{figure}

We now present the event detection results on data from the same date using \textbf{MED}. Table~\ref{tab:wavelet_clustersinfo} summarizes the top 10 clusters detected by \textbf{MED}, four of which are visualized on the map in Fig.~\ref{fig:wavelet_clusters}. From Fig.~\ref{fig:wavelet_clusters} and the third column of Table~\ref{tab:wavelet_clustersinfo}, we see that \textbf{MED} is able to detect events that spread in much larger spatial areas or longer time intervals than \textbf{LED}. Specifically, we see in Fig.~\ref{fig:wavelet_clusters}(a) and Fig.~\ref{fig:wavelet_clusters}(b) two clusters related to OWS protests at Zuccotti Park (cluster 1), and Union Square and Foley Square (cluster 2), respectively, both of which span rather long time intervals. Moreover, although most of the tweets in the two clusters are mainly posted from locations where the protests took place, there also exist tweets in the clusters that mention the same events but have been posted at quite distant locations. In Fig.~\ref{fig:wavelet_clusters}(c) and Fig.~\ref{fig:wavelet_clusters}(d), we see two clusters corresponding to the Raise Cache tech event (cluster 5) and the Mastercard free lunch promotion event (cluster 9), respectively, both of which are more concentrated in time but spread in space (with a few outliers in the latter case). Although there exists certain amount of noise tweets in the detected clusters, these examples demonstrate that \textbf{MED} is able to detect events that concentrate only in time or space, many of which are of different scales. In comparison, \textbf{LED} is not able to detect such event clusters. Specifically, \textbf{LED} produced many separated clusters for OWS protests, two separated clusters with some missing tweets for the Raise Cache tech event, and missed completely the Mastercard promotion event due to the lack of a group of tweets that are concentrated in both time and space.

Finally, we notice that even in the results obtained by \textbf{MED} there sometimes exists more than one cluster about the same event, for example, in Table~\ref{tab:wavelet_clustersinfo} there are two clusters detected for both the OWS protests (clusters 1 and 2) and the Katy Perry concert (clusters 3 and 4). 
First, the protests at Zuccotti Park took place from the morning to noon, while the protests at Union Square and Foley Square happened in the afternoon after 3pm. Although there indeed exist semantic links between tweets that correspond to these two events, the rather different locations and timestamps lead to separate clusters. Second, for the Katy Perry concert, the two clusters highly overlap in both time and space, and the tweets in one cluster have quite strong links to those in the other one. In this case, clusters have been separated mainly because of the strong patterns present in the texts: While in cluster 3 the concert is described mostly using a single term ``katyperry'', in cluster 4 we see two separate terms ``katy'' and ``perry''.

\subsection{Scalability}
The computational complexity of both \textbf{LED} and \textbf{MED} mainly depend on (i) the construction of a similarity graph, and (ii) the graph-based clustering process. As we mentioned before, the Louvain method used in the clustering process is empirically observed to be able to scale to large scale graphs. Therefore, we mainly discuss the computational cost of constructing similarity graphs in the two algorithms.

For both \textbf{LED} and \textbf{MED}, the construction of a similarity graph can be performed efficiently because the similarities need to be computed only for pairs of tweets that have common terms. Thus, the computational complexity of the similarity graph construction, using an appropriate index structure (such as an inverted index), can be $O(n\times \text{avg}\_\text{connectivity})$, where $n$ is the total number of tweets and $\text{avg}\_\text{connectivity}$ denotes, given a tweet $t$, the average number of tweets in the dataset with non-zero similarity with $t$. In our real world experiment, $\text{avg}\_\text{connectivity}$ corresponds to only 2\% of the total number of tweets.

In addition, $\text{avg}\_\text{connectivity}$ can be further reduced by the term-filtering procedure that is employed in \textbf{MED} for noise-filtering. Since term-filtering is applied to the most popular (frequent) terms, this can substantially affect $\text{avg}\_\text{connectivity}$. In our experiment, for example, after the filtering procedure $\text{avg}\_\text{connectivity}$ is further reduced by more than 40\% compared to \textbf{LED}. Moreover, the filtering procedure potentially represents a tradeoff between the performance of the algorithm and its computational complexity. A more aggressive filtering can largely attenuate the influence of noisy information and at the same time reduce the computational cost. However, it might also filter out terms that are related to some relatively small-scaled events.

For \textbf{MED}, we need to compute the spatiotemporal similarity of time series for the valid terms (after term-filtering) shared by every pair of tweets. However, since the spatiotemporal similarity is defined between time series that come from different geographical cells, we only need to evaluate, for each valid term, the pairwise similarity between time series from different cells, instead of comparing every pair of different tweets containing that term. This keeps the number of DWT computations needed relatively low due to the small number of geographical cells.

Practically, for the daily Twitter stream with geotag in the middle and lower Manhattan area of New York City that we have considered in the experiment ($\sim$8000 geotagged tweets with 36000 terms in total), it takes only a few seconds to finish the construction of the similarity graph in \textbf{LED}. For the implementation of \textbf{MED}, it takes roughly 5 minutes for our MATLAB code to create the similarity graph on a lab server with average computing power or 8 minutes on a mid-2009 MacBook Pro (both single core process), where the main computational cost is due to the DWT computations. While we consider this computation time reasonable given the benefits of the algorithm, we certainly hope to further improve the scalability of our algorithm in future work.

\section{Related work}
\label{sec:literature}

Social media data have become pervasive due to the fast development of online social networks since the last decade. This has given rise to a series of interesting research problems such as event detection based on user-generated content \citep{Sayyadi09,Becker09,Aggarwal12}. As an example, \citet{Chen09} and \citet{Papadopoulos11} have proposed to detect social events using tagged photos in Flickr. A more popular platform is Twitter, which has attracted a significant amount of interest due to the rich user-generated text data that can be used for event detection \citep{Atefeh13}. Early works in the field have focused on more specific types of events, such as news \citep{Sankaranarayanan09} and earthquakes \citep{Sakaki10}, while recent approaches detect various types of events \citep{Petrovic10,Marcus11,Becker11,Ozdikis12,Li12b,Parikh13,Berlingerio13}. Although the specific techniques presented in the state-of-the-art event detection approaches may vary from a technical point of view, many of them rely on the detection of certain behaviors in the Twitter stream such as the burstiness of certain keywords, which indicates the emergence of particular events. In particular, several works use wavelets, which is a well-developed tool in signal processing, for event detection based on keyword burstiness patterns \citep{Weng11,Cordeiro12}.

Recently, there has been an increasing amount of interest in exploring both the temporal and spatial dimensions to better capture the meaningful information and reduce noise in the data from social media platforms. In \citet{Rattenbury07}, the authors have proposed to analyze for event extraction the semantics of tags associated with the Flickr photos, by taking into account multiple temporal and spatial resolutions. In \citet{Chen09}, the authors have proposed to cluster Flickr photos based on both the temporal and the spatial distributions of the photo tags using wavelets. In \citet{Becker10}, the authors have considered combining text, temporal and spatial features in order to build an appropriate tweet similarity measure. In \citet{Lappas12}, the authors have proposed two approaches to detect burstiness of keywords in both temporal and spatial dimensions simultaneously. In \citet{Sugitani13}, the authors have proposed a hierarchical clustering procedure for event detection in Twitter, where both temporal and spatial constraints have been imposed to measure the similarities of tweets. They have also proposed to examine co-occurrences of keywords that present specific spatiotemporal patterns. Other examples include \citet{Lee11,Li12a,Thom12,Walther13} and \citet{Zaharieva13}, where the authors have proposed spatiotemporal clustering methods for anomaly and event detection in Twitter and Flickr, respectively. These approaches are certainly inspirational to the idea proposed in the present paper; However, most of them do not explicitly handle multiple spatiotemporal scales in event detection.

Finally, there are a few approaches in the literature that have studied the influence of different resolutions for temporal and spatial analysis in event detection. For example, in \citet{Cooper05} and \citet{Rattenbury07}, the authors have proposed to use a scale-space analysis of the data \citep{Witkin83}. The common objective in these approaches is to select the most appropriate scale for event extraction and detection. More generally, multiscale or multiresolution clustering algorithms has been of interest in the machine learning, pattern recognition, and physics \citep{Ronhovde11,Ronhovde12} communities since the last decade. The approaches that take advantage of the properties of the wavelet transform to enable a multiresolution interpretation in the clustering process, such as the works in \citet{Sheikholeslami00} and \citet{Tremblay12}, are of particular interest. Although these approaches are not originally proposed for event detection in social media platforms, they have inspired us to consider wavelets in our framework. While they output multiple sets of clustering solutions at different resolutions, our approach however uses wavelets to choose the appropriate temporal and spatial resolutions for constructing a single data similarity graph.

In summary, although there exist many approaches that take into account the temporal and spatial dimensions of the social media data for event detection, they generally do not explicitly handle different scales in data analysis. In contrast, our framework explicitly handles multiple spatiotemporal scales, which we believe is essential for building an efficient and generic event detection approach. Different scales in the temporal and spatial dimensions have been treated separately in most of the state-of-the-art analyses, but the relationship and interaction between these scales have been largely overlooked in the literature. To the best of our knowledge, our approach is the first attempt that is based on an explicit modeling of the relationship between different temporal and spatial resolutions. Finally, we present a statistical analysis of the temporal and spatial distributions of noisy information in the Twitter data, which we believe is the first of its kind. We believe our perspective contributes to the research in the field of social media analytics and provides new insights into the design of novel clustering and event detection algorithms.

\section{Conclusion}
\label{sec:conclusion}
In this paper, we have proposed a novel approach towards multiscale event detection in social media. Especially, we have shown that it is important to understand and model the relationship between the temporal and spatial scales, so that events of different scales can be separated simultaneously and in a meaningful way. Furthermore, we have presented statistical modeling and analysis about the spatiotemporal distributions of noisy information in the Twitter stream, which not only helps us define a novel term-filtering procedure for the proposed approach, but also provides new insights into the understanding of the influence of noise in the design of event detection algorithms.
Future directions include (i) further investigation of the possibility of extending and generalizing the proposed scale relationship model to handle temporal and spatial scales simutaneously for multiscale event detection, (ii) more appropriate and accurate statistical models for analyzing noisy information present in social media data, and (iii) improvement on the scalability of the proposed algorithms.


\bibliographystyle{spbasic}      
\bibliography{mybibfile}   

\end{document}